\documentclass{aa}  

\usepackage{graphicx}

\usepackage{txfonts}

\usepackage{natbib}
\bibpunct{(}{)}{;}{a}{}{,}

\usepackage{xcolor}

\usepackage[switch]{lineno}

\begin{document}

   \title{Compact groups of galaxies in the TNG100 simulation}

   \author{Bruno M. Celiz
          \inst{1,2,3}\fnmsep\thanks{\email{bruno.celiz@mi.unc.edu.ar}}
          \and
          José A. Benavides\inst{4}
          \and
          Mario G. Abadi\inst{2,3}
          }

    \institute{Facultad de Matemática, Astronomía, Física y Computación, UNC, Medina Allende s/n, X5000HUA, Córdoba, Argentina
        \and
            Instituto de Astronomía Teórica y Experimental, CONICET--UNC, Laprida 854, X5000BGR, Córdoba, Argentina
        \and
            Observatorio Astronómico de Córdoba, UNC, Laprida 854, X5000BGR, Córdoba, Argentina
        \and
            Department of Physics and Astronomy, University of California, Riverside, CA, 92521, USA}

   \date{Received XXX; accepted YYY}
 
  \abstract
  {Using the TNG100 cosmological simulation, we study the formation and evolution of compact groups of galaxies. Over a redshift range of $0 \lesssim z \lesssim 0.2$, we identify these compact groups as FoF galaxy groups with high mean surface brightness ($\overline{\mu}_r < 26.33 ~ \mathrm{mag~arcsec^{-2}}$) and a minimum of 4 galaxy members.
  Typically, our compact groups have a median characteristic size of $\sim$$150$ kpc, 1D velocity dispersions of $150 ~ {\rm km ~ s^{-1}}$, and stellar masses around $2\times 10^{11} ~ M_{\odot}$. Roughly 1\% of galaxies of stellar mass above $10^9 ~ M_{\odot}$ lie in physically dense compact groups. We found that these systems do not constitute a separate category within the broader population of galaxy groups; instead, they represent the lower end of the size distribution in the sequence of galaxy group sizes. We traced their evolution backward in time, revealing that they initially form as galaxies systems with a mean low surface brightness that systematically increases to a peak value before stabilizing over time, exhibiting oscillatory behaviour over the following several Gyrs during which mergers may occur. Mergers often transform compact groups with typically four members into galaxy pairs or triplets, which may eventually can increase again their number of members accreting a new galaxy. Nevertheless, the full merging of all constituent galaxies into a single massive galaxy is a rare phenomenon.}

   \keywords{
   galaxies: groups: general --
   galaxies: interactions --
   galaxies: kinematics and dynamics
    }

   \maketitle

\section{Introduction} \label{sec:intro}

Compact groups (CGs) are selected as isolated systems of a few galaxies ($N \sim 4$) with similar apparent magnitude ($\Delta m < 3$ mag) separated by distances of a few tens of kpc. They exhibit extremely high mean surface brightnesses ($\mu < 26 ~ \mathrm{mag~arcsec}^{-2}$), comparable to the cores of rich galaxy clusters \citep[see e.g.,][]{Hickson1992,McC2009,DG2012,Sohn2015}. Their high densities, combined with relatively low velocity dispersions ($\sigma_v \sim 300~\mathrm{km~s}^{-1}$), result in estimated crossing times of only a few hundredths of the Hubble time \citep{Mamon1986,DG2020,Zheng2021}. Consequently, the member galaxies are expected to merge within 1-2 Gyr \citep{Carnevali1981,Mamon1987,Governato1991,McC2008}, making CGs ideal laboratories for studying strong galaxy interactions and their impact on galactic evolution.

The existence of systems with such extreme apparent compactnesses is not well understood. Pioneer works theorised about the nature of CGs, arguing that they are physically dense configurations of galaxies \citep{HicksonRood1988}, or be transient configurations within looser (non-compact) groups \citep{Rose1979}. However, CGs can also be produced by chance alignment of galaxies within non-compact groups \citep{Rose1977, Mamon1986,WalkeMamon1989,Mamon2008} or even within filaments \citep{Hernquist1995}.

Studies of compact group abundances at redshifts $0 < z < 0.2$, conducted in large surveys like the Sloan Digital Sky Survey \citep[SDSS,][]{York2000}, demonstrate their presence at different cosmic times \citep[see e.g.,][]{Sohn2015,DG2018,Zheng2021,Zheng2022}. Several theories have been proposed to reconcile the observed abundances of CGs with their seemingly short lifetimes, e.g. if mergers occur rapidly, new CGs must continually form to maintain the observed space density.

Regarding the evolution of physically dense CGs, \cite{Diaf1994} suggested that a high accretion rate of new galaxies, comparable to the expected merger rate, could allow a constant replenish of galaxies within existing groups. Alternatively, CGs might form from looser (non-compact) galaxy groups through angular momentum loss due to dynamical friction \citep{Schneider1982}. Another possibility is that many identified CGs are fortuitous alignments of galaxies in configurations elongated along the line of sight, resulting in regions of of high surface brightness due to projection effects \citep{Mamon1986}. Finally, if CGs do survive longer than expected, they should maintain their dense, compact configurations for several Gyr \citep{DG2021}.

Large cosmological simulations enable statistical analysis of CGs properties and address issues of abundance and projection effects by providing 3D galaxy positions. Several authors analysed semi-analytical models of galaxy formation ran on gravity-only simulations to create mock galaxy catalogues where CGs can be identified, measuring the fraction of CGs that are physically compact and how many are identified as such only due to a fortuitous alignment of the members \citep[see e.g.,][]{McC2008, DG2010}. According to the distance between galaxies or the sphericity of groups \citep[see][respectively]{McC2008, DG2010}, 30-70\% of CGs identified using observational criteria are physically compact. More recently, \citet{DG2020} show that the contamination of CGs by chance alignments of galaxies is $\sim$50\%, using newer semi-analytical models on cosmological simulations with more realistic initial cosmological parameters.

CGs, on the other hand, can also be identified using 3D search algorithms \citep[see e.g.][]{Taverna2016}. \citet{Taverna2022} found that $\sim$90\% of all CGs identified using Hickson criteria are substructures within large groups or clusters. These physically dense systems are found throughout cosmic time, and a significant fraction of galaxies have been part of such systems at a certain point of their evolution \citep{Wiens2019}.

Early studies of compact group evolution used numerical methods to examine the survival of individual CGs under various idealised initial conditions and dark matter distributions \citep[see e.g.,][]{Carnevali1981,Mamon1987,Barnes1989,Governato1991,Athanassoula1997}. More recently, these systems have been analysed within realistic cosmological contexts using gravity-only simulations \citep[see e.g.,][]{McC2008, Farhang2017,Tza2019,DG2021} and hydrodynamical simulations, which incorporate baryons within large volumes and offer a complementary approach to semi-analytical models applied to gravity-only simulations. For example, \citet{HyP2020} used data from the Evolution and Assembly of GaLaxies and their Environments cosmological hydrodynamical simulation \citep[EAGLE\footnote{\url{https://icc.dur.ac.uk/Eagle/}},][]{EAGLE2015} to characterize CGs identified in a flux-limited mock catalogue. While the former approach isolates group evolution based on pre-defined parameters, the latter provides insights into evolution within more complex environments. Due to the diverse numerical methods employed and the highly non-linear nature of these systems, coalescence times (the timescale over which all members of a group sequentially merge into a single massive galaxy) vary widely, from as short as 0.5 Gyr to as long as 9 Gyr. Consequently, the temporal evolution of CGs is not yet fully understood.

The existence of these physically compact configurations of galaxies entails an open problem for the standard $\Lambda$CDM cosmology. How can observed CG abundances at different redshifts be explained? What role do they play in hierarchical structure formation? Are they a special kind of groups of galaxies or a mere transient configuration of non-compact groups? How long do CGs persist? While state-of-the-art cosmological hydrodynamic simulations emerged as powerful tools for modelling galaxy formation and evolution in the past decade, their full potential remains to be explored in the context of compact group studies. \citet{HyP2020} found that the space density of simulated CGs remains almost constant at redshifts $0<z<0.2$, comparing it with that of CGs extracted from the Sloan Digital Sky Survey (SDSS) Data Release 6 by \cite{Sohn2015}. The balance between CG formation and destruction can be explained considering that the majority of simulated CGs ($\sim$$60$\%) are transient configuration of groups of galaxies, only compact in projection, while those that are physically dense show typical coalescence times of 2-3 Gyr. Aiming to further characterize the dynamical evolution of physically compact groups, elucidating the mechanisms that cause a system to become or no longer be a CG, in this work we use the high-resolution cosmological hydrodynamic simulation from the IllustrisTNG suite, TNG100-1, to analyse the past and future evolution of CGs. This simulation has similar characteristics than the previously mentioned EAGLE simulation, but the scope of this work is to focus on the compaction time and occurrence of mergers or accretion of massive galaxies within CGs, defined as isolated groups of four or more similar mass galaxies in physically compact configurations, hence identifying them in 3D space.

This paper is organised as follows: in Sect. \ref{sec:methods} we present details of the simulation used and the CGs identification criteria. In Sect. \ref{sec:CGs_z0} we define our samples, compute the abundances of these systems compared to other works, analyse the sample of CGs at redshift $z=0$ and take a glance at its backward temporal evolution. In Sect. \ref{sec:CG_z-geq-0} we study the evolution of CGs formed before or
at $z=0$, their survival as CGs, their galaxy merger rates and their possible full coalescence. We briefly discuss our results in Sect. \ref{sec:discussion}, and finally in Sect. \ref{sec:conclusion} we present our conclusions.

\section{Methods} \label{sec:methods}

\subsection{Simulation} \label{sec:simus}

We identify and follow the evolution of CGs in The Next Generation Illustris Simulations \citep[IllustrisTNG\footnote{\url{https://www.tng-project.org/}},][]{Miranacci2018, Naiman2018, Pill2018,Springel2018,Nelson2019}, a suite of magneto-hydrodynamic cosmological simulations of a standard $\Lambda$CDM Universe \citep[$h = 0.6774;~\Omega_m = 0.3089;~\sigma_8 = 0.8159$, consistent with][]{Planck2016}. The initial conditions are set at redshift $z=127$ have been obtained
with the code \texttt{N-GENIC} \citep{Springel2005}, and the simulations are evolved with the moving mesh code \texttt{AREPO} \citep{Springel2010, Pakmor2016} forward in time until $z=0$. The properties of dark matter and baryon particles are systematically recorded in 100 snapshots at intervals of $\sim$0.15 Gyr. From this set of simulations we use the data from TNG100-1, the highest resolution run of the $110.7^3$ Mpc$^3$ box (hereafter TNG100). This simulation contains $1820^3$ dark matter particles of mass $m_{\mathrm{DM}} = 7.5 \times 10^6~\mathrm{M_{\odot}}$, and an equal number of initial gas cells, with a ``target baryon mass" of $m_{\mathrm{baryon}} = 1.4 \times 10^6~\mathrm{M_{\odot}}$, adopting the same force softening length for dark matter and star particles $\epsilon = 0.74$ kpc (at redshift $z=0$). TNG100's baryonic treatment \citep[][]{Weinberger2017,Pillepich2018} is an update from that of the Illustris project \citep[][]{Vogelsberger2013,Vogelsberger2014}, assuming a Chabrier initial mass function \citep{Chabrier2003} and modelling stellar evolution as in \citet{Pillepich2018}.

We use the halo and subhalo catalogues available for each snapshot, generated using a \texttt{friends-of-friends} with a particle linking length parameter $b = 0.2$ \citep[FoF,][]{Davies1985} and the \texttt{SUBFIND} \citep{Springel2001, Dolag2009} algorithms. We use the \texttt{SubLink} \citep{R-G2015} merger tree to track the temporal evolution of identified galaxies. We consider galaxies with $M_* > 10^9 ~ \mathrm{M_{\odot}}$, ensuring well-resolved subhalos with more than $\sim$1000 stellar particles within twice the half-mass radius (using \texttt{SubhaloMassInRadType[4]} from \texttt{SUBFIND}). Position, velocity, stellar mass, SDSS $r$-band luminosity, and group membership are extracted from each snapshot's galaxy catalogues. All sizes are converted from comoving to physical kpc.

\subsection{Identification criteria} \label{sec:criteria}

Although there is no unique definition of a CG, in observational works galaxy group compactness is typically quantified by mean surface brightness $\mu$ \citep[see e.g.,][]{McC2009, DG2012, Sohn2015}, defined as follows:
\begin{equation}
    \mu = -2.5 \log \left( \frac{\sum_{i=1}^N 10^{-0.4 m_i}}{\pi ~ \theta^2} \right)
    \label{eq:mu_r}
\end{equation}
where $m_i$ is the apparent magnitude of the $i$th galaxy member of the CG and $\theta$ is the angular radii of the smallest circle enclosing all members. Using TNG100 data, we calculate surface brightness in terms of luminosity $L_i$ and the projected radius $R_{\rm p}$, then convert from $\mathrm{L_{\odot}~kpc^{-2}}$ to $\mathrm{mag~arcsec^{-2}}$ using:
\begin{equation}
    \mu = 36.57 -2.5 \log \left( \frac{\sum_{i=1}^N L_i/\mathrm{L}_{\odot}}{\pi ~ R^2_{\rm p}/\mathrm{kpc^2}} \right) + \mathcal{M_{\odot}}
    \label{eq:mu_right}
\end{equation}
where $\mathcal{M_{\odot}} = 4.65$ is the absolute magnitude of the Sun in the SDSS $r$-band \citep{Willmer2018} and the projected radius is the maximum projected distance from the centroid to a member of the group. Lower $\mu$ values indicate higher surface brightness, and thus, more compact groups.

To investigate physically compact groups of galaxies, we define them as high mean surface brightness groups, independent of projection, consisting of four or more galaxies with similar stellar masses. Our CG identification criteria are specifically designed to select for intrinsically compact configurations, allowing us to directly study the inherent properties and evolution of these systems. Our CG identification criteria are:

\begin{itemize}
    \item Isolation: The CG members perfectly match the galaxies of the single host group with stellar masses above one-tenth of the central ($M_{*,i} > M_{*,\mathrm{central}}/10$).
    \item Richness: They contain at least 4 members ($N \geq 4$). 
    \item Compactness: The geometric mean surface brightness of the CG, measured along the three cartesian axes, must be above a threshold corresponding to $\overline{\mu}_r < 26.33 ~ \mathrm{mag~arcsec^{-2}}$
\end{itemize}
where $M_{*,\mathrm{central}}$ is the central galaxy's stellar mass, $M_{*,i}$ is the $i$th member's stellar mass, $N$ is the number of galaxies with stellar mass similar to the central in the FoF Group (including the central), and $\overline{\mu}_r$ is the mean surface brightness (Equation \ref{eq:mu_right}) averaged over three orthogonal spatial projections: $\overline{\mu}_r = \left(\mu_{XY} + \mu_{XZ} + \mu_{YZ}\right)/3$. This average provides a projection-independent compactness estimator, which is measured only accounting for the $N$ galaxy members. The $\mu$ threshold value used to classify a group as compact varies between works and depends on the photometric band utilized. Since we use mock absolute magnitudes in the SDSS $r$-band, we follow \cite{Taverna2016} and adopt $\overline{\mu}_r = 26.33 ~ \mathrm{mag~arcsec^{-2}}$ as our CG identification threshold. The arithmetic mean of the three cartesian surface brightnesses, expressed in surface magnitudes, provides a good measure of physical density of CGs, since if a CG is a chance alignment of galaxies along the line of sight, it will usually appear compact in only one of three cartesian lines of sight and the mean surface brightness is likely to fail to meet the compactness threshold.

We note that our CGs always include the central galaxy of a FoF group (by definition) and thus cannot be substructures of larger groups. By analysing truly isolated groups, we avoid including a large number of CGs-within-groups, which \citet{Taverna2022} reported to be $\sim$90\% of all redshift-space identified systems.

\begin{figure}
    \includegraphics[width=\columnwidth]{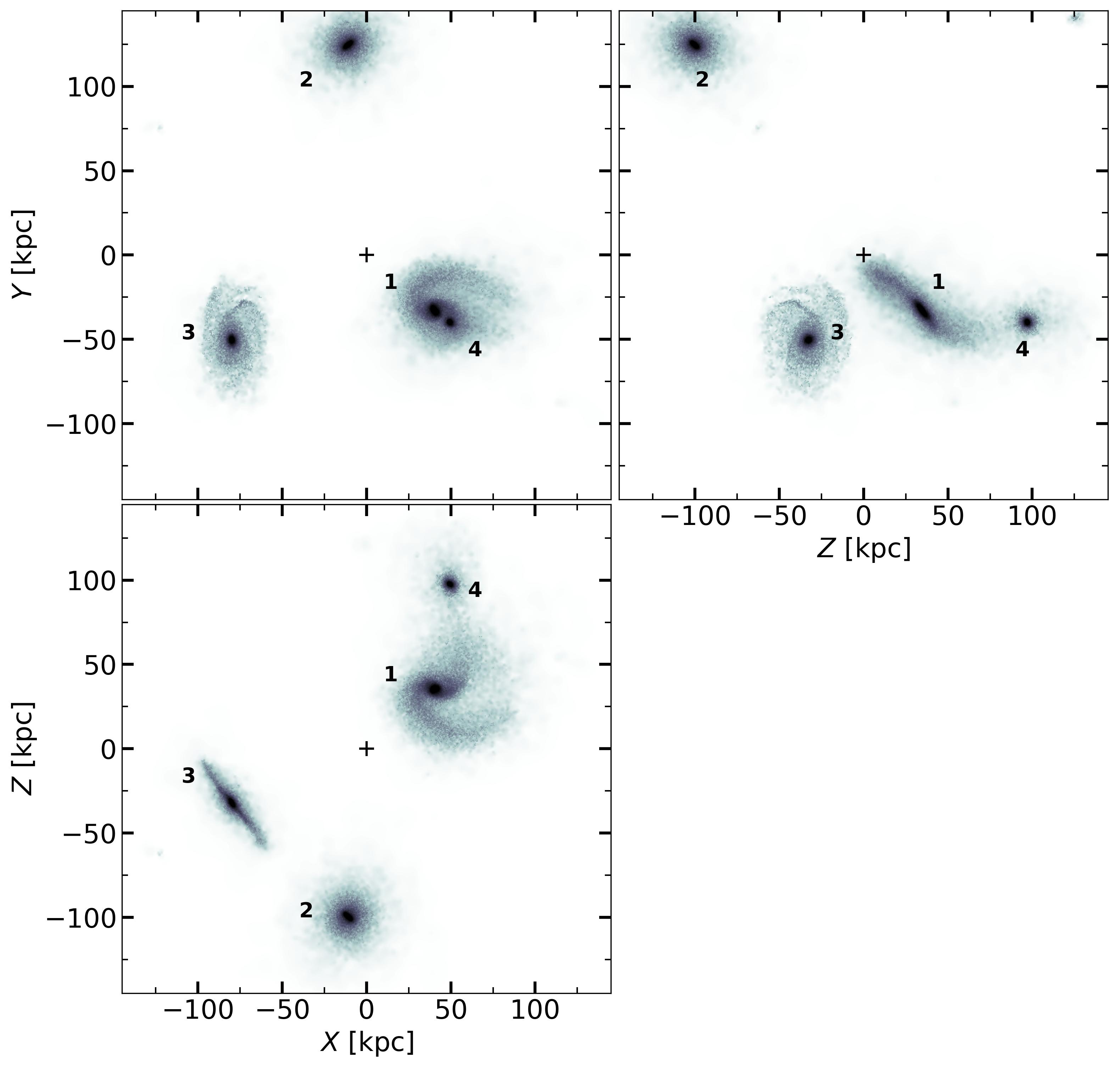}
    
    \caption{Spatial distribution of stellar particles in three orthogonal projections ($XY, XZ, YZ$) for a redshift $z=0$ CG. This group (\texttt{FoF group \#469}) consists of four galaxies with similar stellar mass in a physically compact configuration ($\overline{\mu}_r = 25.83~\mathrm{mag~arcsec^{-2}}$). The mean projected radius is $\overline{R}_{\rm p} = 131.34$ kpc, with total stellar mass $M_{*,T} = 10^{11.0}~ M\mathrm{_{\odot}}$ and a 1D-velocity dispersion $\sigma_{v,\rm 1D} = 109.96~\mathrm{km~s^{-1}}$ (see Table \ref{tab:CGs_z0}). Members are labelled according to their stellar mass, being 1 the central and more massive of the system. 1 and 4 show strong tidal interaction, while the other two have disc-like and spheroidal morphologies (3 and 2, respectively). Darker colours indicate a higher number density of particles. Images generated with the Py-SPH Viewer library \citep{BenitezLlambay2017}.}
    \label{fig:sph_469}
\end{figure}

Fig. \ref{fig:sph_469} illustrates a simulated CG identified with this criteria and shows the projected spatial distribution of stellar particles of a CG in 3 orthogonal projections $(XY,XZ,YZ)$ within $200 \times 200~\mathrm{kpc^2}$ regions, centred on the group's geometric centre. This group consist in 4 galaxies with stellar mass $M_{*,i} > 10^{10}~M\mathrm{_{\odot}}$, conforming a physically compact system of $\overline{\mu}_r = 25.83~\mathrm{mag~arcsec^{-2}}$. The properties of this CG are highlighted in Table \ref{tab:CGs_z0}. Members are ranked according to their stellar mass, and exhibit diverse morphologies. While 3 shows a disc morphology, 2 is a spheroid, and tidal features emerge between 1 and 4, highlighting the strong interactions within these environments.

\section{Compact groups in TNG100} \label{sec:CGs_z0}

\subsection{Compact groups at $z=0$} \label{sec:CG_z0}

\begin{figure*}
    \centering
    \includegraphics[width=\textwidth]{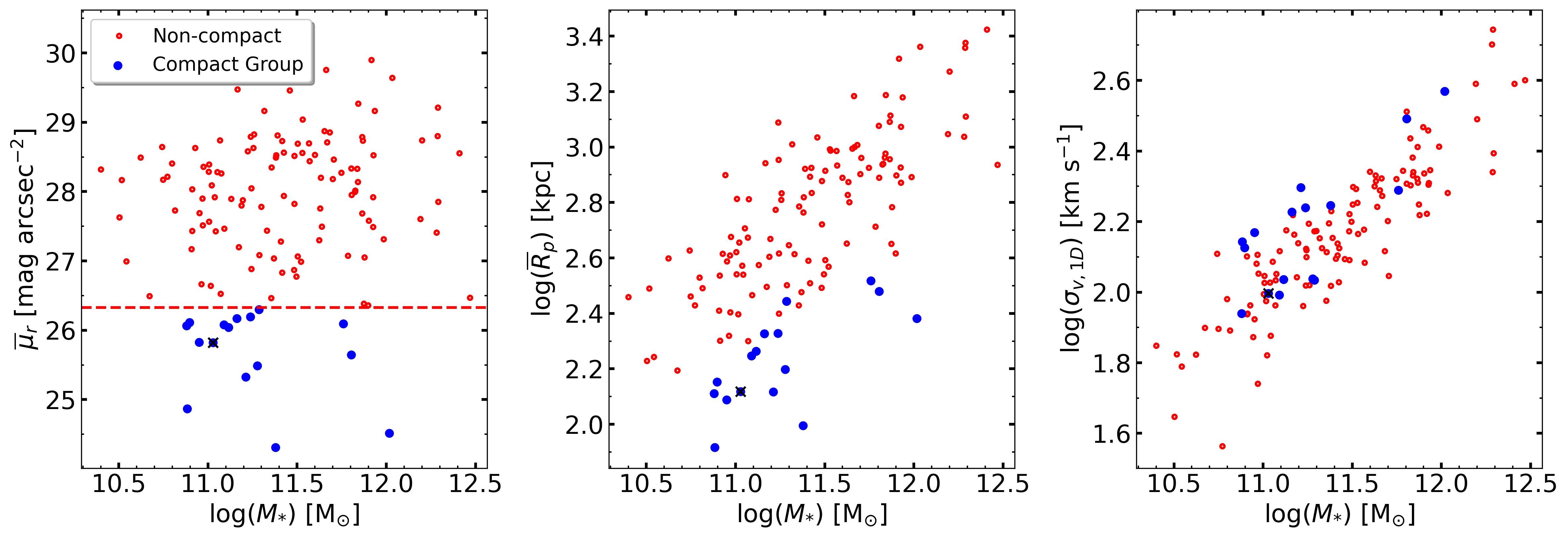}
    \caption{Redshift zero correlations of CG properties with total group stellar mass. From left to right: i) Mean surface brightness $\overline{\mu}_r$; ii) mean projected radius $\overline{R}_{\rm p}$; and iii) 1D-velocity dispersion $\sigma_{v,\rm 1D}$ for all groups with four or more galaxies in TNG100. The threshold value $\overline{\mu}_r = 26.33~\mathrm{mag~arcsec^{-2}}$ (horizontal dashed red line, left panel) defines the sample of CGs (blue filled circles), while non-compact groups (red empty circles) lie above this threshold. CGs are the smallest groups at a given stellar mass, but their velocity dispersions are similar to those of non-compact groups. The values of the CG of Fig. \ref{fig:sph_469} are shown with black crosses.}
    \label{fig:mu-r_p-sigma-M_T_z0}
\end{figure*}

Applying our criteria to the TNG100 halo and subhalo catalogues at redshift $z=0$ (cosmic time $t_{\mathrm{cosmic}} \approx 13.8$ Gyr), we identified 131 groups with four or more similar-mass galaxies. Among these groups, 15 ($\sim$11\%) satisfy $\overline{\mu}_r < 26.33 ~ \mathrm{mag~arcsec^{-2}}$, constituting our $z=0$ CG sample. This implies that $\sim$0.94\% of all galaxies with $M_* > 10^9 ~M_{\odot}$ are members of CGs, fairly in agreement with the fraction reported by \citet{Wiens2019}.

In order to asses the projection effects on the identification of CGs, we compare the amount of systems that satisfy the compactness criterion over random projections to the amount of true CGs that are found. 60-70\% of CGs identified in projection are truly compact with our criteria, a slightly higher fraction than that found by \citet{DG2020}.

In Fig. \ref{fig:mu-r_p-sigma-M_T_z0} we show properties of all groups\footnote{We only account for galaxies with stellar mass similar above one-tenth of the central of a group to measure these properties.}, compact or not, as function of the total stellar mass. The threshold value $\overline{\mu}_r = 26.33~\mathrm{mag~arcsec^{-2}}$ separates CGs from non-compact groups. With black crosses we indicate the values of the illustrative example shown in Fig. \ref{fig:sph_469}. These CGs have 4, 5 or 7 members\footnote{There are not CGs of 6 galaxies in this sample.} (12, 2 and 1 CGs, respectively), and have sizes ranging between 80-320 kpc, systematically smaller than those non-compact at fixed total stellar mass, as expected (see Equation \ref{eq:mu_right}). However, this does not hold for CGs with stellar mass to light ratios $M\mathrm{_{*}}/L_{r} < 1$ in solar units, allowing larger sizes for those CGs with higher luminosities. On the other hand, there is no segregation between compact and non-compact groups regarding their velocity dispersion. The broad range of velocity dispersions ($\sigma_{v,\rm 1D} =$ 90-400 $\mathrm{km~s^{-1}}$) is directly related with the total stellar mass of CGs, spanning over an order of magnitude ($M_{*} = 10^{10.8\mathrm{-}12.2}~M\mathrm{_{\odot}}$).

These velocity dispersions are roughly half of the virial velocity $v_{200,c} = \sqrt{G~M_{200,c}/R_{200,c}}$, where $R_{200,c}$ is the radius of the sphere of overdensity 200 times the critical density of the Universe, while $M_{200,c}$ is the mass within $R_{200,c}$. The factor 0.5 is well below the factor $\sim$0.68 predicted for number-follows-mass NFW models for groups \citep[eq. B6 of][]{Mamon2013}, and also lower than the factor $\sim$0.62 deduced from the analysis of velocity bias of galaxies in clusters by \citet{Anbajagane2022}. We show $\sigma_{v,\rm 1D}$ as function of $v_{200,c}$ for compact and non-compact groups in Fig. \ref{fig:App_sigma_v200}. Since galaxy mergers are more frequent in low-mass groups compared to clusters \citep{Mamon1992,KontorovichKrivitsky1997,MakinoHut1997}, this suggests that the low velocity dispersions of low-mass groups, given their total mass, are caused by galaxy mergers that transform orbital energies into the internal energy of the remnant.

These results suggest that CGs are not statistically different systems of galaxies, but rather represent the high-end of the surface brightness distribution of all groups, in agreement with recent observational works that compare CGs to non-compact groups \citep{Zheng2022,Tricottet2025}. Hence, the $\overline{\mu}_r = 26.33~\mathrm{mag~arcsec^{-2}}$ threshold is somewhat arbitrary, simply defining the CG sample size. The typical sizes and velocity dispersions of our CGs indicate that the crossing time of galaxies within these systems is 0.4-1.0 Gyr. Properties of all $z=0$ CGs are listed in Table \ref{tab:CGs_z0}.

\subsection{Assembly of compact groups} \label{sec:assembly_CG}

\begin{figure*}
    \includegraphics[width=\textwidth]{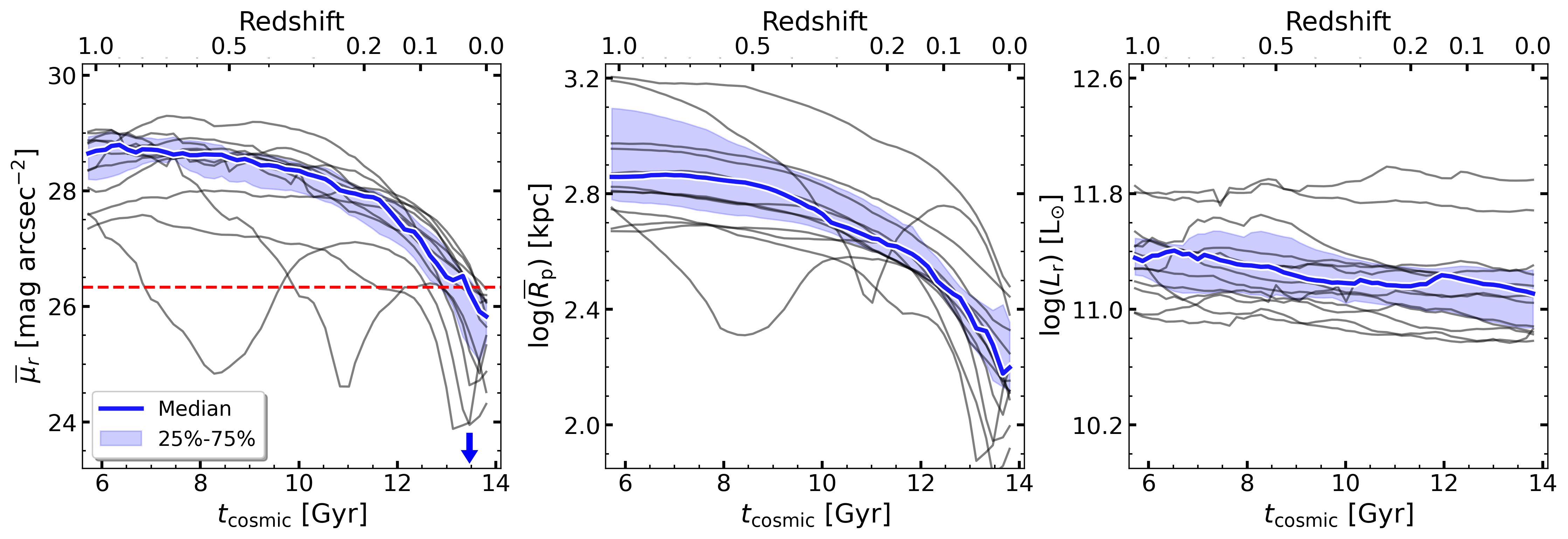}
    \caption{Evolution of CG properties. From left to right: i) group surface brightness $\overline{\mu}_r$; ii) size $\overline{R}_{\rm p}$; and iii) total luminosity $L_{r}$. Solid blue lines show the sample medians, and shaded regions encompass the 25th-75th percentiles. At the left panel, we show the threshold value $\overline{\mu}_r = 26.33~\mathrm{mag~arcsec^{-2}}$ with an horizontal dashed red line, and the median threshold time $t_{\mathrm{thr}}$ is represented with a blue arrow, indicating that more than 50\% of the CGs became compact less than 1 Gyr ago. The other two panels show that surface brightness strongly correlates with size evolution, while luminosity mildly changes ($\Delta \sim 0.2$ dex).}
    \label{fig:mu-t_z0}
\end{figure*}

CGs stand out as groups of $4 \leq N \leq 7$ similar-mass galaxies in physically compact configurations, albeit their velocity dispersions are comparable to those of non-compact groups. To investigate the pathways that lead to CG formation, we use each galaxy's main progenitor to track the CG's mean surface brightness evolution throughout cosmic time, covering $5.9 \leq t\mathrm{_{cosmic}/Gyr} \leq 13.8$ (corresponding to a redshift range $0 \leq z \leq 1$). Mean surface brightness $\overline{\mu}_r$ serves as a measure of group compactness, proportional to the galaxy luminosities and inversely proportional to the square of the group's projected radius in three orthogonal directions, as shown in Equation \ref{eq:mu_right}. Therefore, the evolution of a group's surface brightness is driven by both stellar evolution and star formation within galaxies, which affects the total luminosity, and the group's dynamical evolution, which alters member positions and thus the group's size. In Fig. \ref{fig:mu-t_z0} we show the temporal evolution of these properties for each of the 15 CGs identified. We define as "threshold time" ($t_{\mathrm{thr}}$) the earliest cosmic time when the group satisfies the compactness criterion, i.e. when its mean surface brightness $\overline{\mu}_r (t)$ drops below the threshold value $26.33~\mathrm{mag~arcsec^{-2}}$. The median $t_{\mathrm{thr}}$ value ($\sim$13.5 Gyr, blue arrow on the left panel) indicates that most CGs became compact less than 1 Gyr ago, with only 2 groups ($\sim$13\%) exhibiting $t_{\mathrm{thr}} \lesssim 10$ Gyr and the older one showing $t_{\mathrm{thr}} \approx 7$ Gyr. Among the diversity of curves, we note that CGs can become compact for the first time at early ($\sim$7.0 Gyr) or late ($\gtrsim$13.0 Gyr) cosmic times.

The centre and right panels of Fig. \ref{fig:mu-t_z0} show the temporal evolution of the group size and the total luminosity, respectively. In this range of cosmic time, the total luminosity of CGs changes only slightly ($\Delta L \sim -0.2$ dex), while group size decreases significantly ($\Delta R_{\rm p} \sim -0.6$ dex), from $650 \pm 150$ kpc at redshift $z = 1$ to $150 \pm 50$ kpc at $z=0$. Although the decrease in luminosity lowers surface brightness, its median increases, mirroring the trend in group sizes. The median surface brightness change of $\sim$2.5 $\mathrm{mag~arcsec^{-2}}$ is, thus, mainly caused by size rather than group luminosity variations. This implies that previously non-compact groups can get smaller over time, achieving sufficiently high surface brightness to be identified as CGs, after which orbital motion leads to oscillatory surface brightness evolution. On the other hand, the mild decrease in total luminosity may be attributed to passive stellar evolution \citep[see e.g.][]{Zandivarez2023}. 

\section{Compact groups at $z>0$} \label{sec:CG_z-geq-0}

\subsection{Abundance of compact groups} \label{sec:abundaces}

\begin{figure}
    \includegraphics[width=\columnwidth]{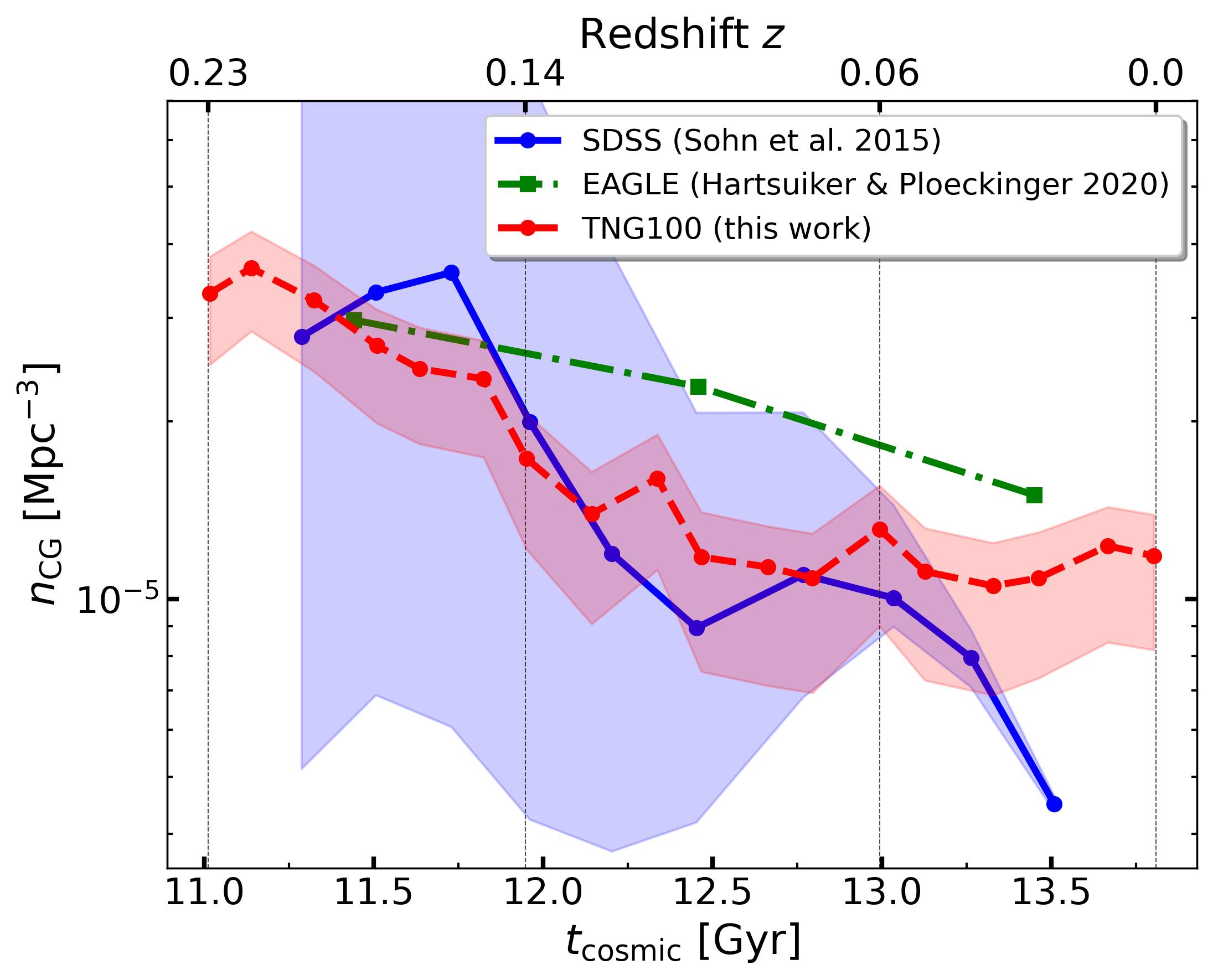}
    \caption{Evolution of space density of CGs obtained in this work (TNG100, dashed red line with circles, shaded regions indicate Poisson errors), by \citet[][mock catalogue from EAGLE, dash-dotted green line with squares]{HyP2020}, and by \citet[][SDSS-DR12, solid blue line with circles]{Sohn2015}. We use vertical dashed black lines to mark four TNG100 snapshots, separated by $\Delta t_{\mathrm{cosmic}} \approx 1$ Gyr. Although at $z \approx 0$ the abundances in TNG100 are a factor $\sim$$2$ higher than the observed in SDSS, at higher redshifts the space density of CGs in TNG100 is consistent with that reported by observational works.}
    \label{fig:dens_num}
\end{figure}

Initial works, starting with \citet{Hickson1992} studied samples of CGs covering redshifts $z < 0.05$. Then, subsequent studies obtained samples of CGs up to $z \lesssim 0.1$ \citep[][with DPOSS II data]{PyI2012}. More recent galaxy surveys, such as SDSS, allow the identification of CGs within the range $0 \lesssim z \lesssim 0.2$ \citep[][]{Sohn2015, Zheng2022}. Since this redshift range is somewhat small, the CG identification criteria used do not change with redshift. We maintain the same criteria described in Sect. \ref{sec:criteria} to identify CGs in 16 snapshots of TNG100, spanning redshifts $0 \leq z \leq 0.23$ (cosmic time $11.0 \lesssim t_{\mathrm{cosmic}}~\mathrm{/Gyr} \leq 13.8$) to compare their space density with observational \citep[][SDSS DR12]{Sohn2015} and theoretical results \citep[][mock galaxy catalogue from the EAGLE simulation]{HyP2020}, as shown in Fig. \ref{fig:dens_num}. The TNG100 CG space density decreases by a factor of $\sim$3 from $t_{\mathrm{cosmic}} \approx 11.0$ Gyr to 12.5 Gyr, remaining almost constantly $\sim$$10^{-5} ~ \mathrm{CGs} ~ \mathrm{Mpc^{-3}}$ until $z=0$. The decrease is larger than the expected for a cosmological expansion-induced decrease $(n_{\rm CG} \propto a^{-3})$ with a constant number of CGs across different redshifts. We identify 23 CGs at $z=0.23$, but only 16 at $z = 0.1$ and 15 at $z=0$. Thus, the apparent flattening (within the Poisson errors, red shaded regions) exhibited at $z < 0.1$ could be caused by chance, where typically less systems are identified as CG at cosmic times $12 < t_{\rm cosmic}/{\rm Gyr} < 13$ and more CG form at $13 < t_{\rm cosmic}/{\rm Gyr}$.

Remarkably, despite comparing CGs identified observationally (SDSS) or in cosmological hydrodynamic simulations (EAGLE and TNG100), the abundances are consistent at redshift 0.1-0.2. We further discuss the discrepancies between works in Sect. \ref{sec:discussion}.

Besides the CG sample obtained at $z=0$ (hereafter CGs$_{z=0}$), we characterize CGs identified within redshifts $0 < z \leq 0.23$. Based on the CGs$_{z=0}$ median threshold time, we avoid sample overlap selecting CGs every 1 Gyr. Specifically, we analyse CGs identified at snapshots 94, 88, and 82, corresponding to redshifts $z=0.06$, 0.14 and 0.23, and lookback
times of 0.81, 1.85, and 2.29 Gyr (shown as vertical lines in Fig. \ref{fig:dens_num}). Combined, these three samples comprise 52 CGs with an amount of galaxy members, sizes, 1-D velocity dispersions, mean surface brightnesses and total stellar masses consistent with the CGs$_{z=0}$ measured properties, i.e, these CGs are systems of 4-7 galaxies of similar stellar mass, with $\overline{R}_p = 80$-500 kpc, $\sigma_{v,\rm 1D} = 50$-500 $\mathrm{km~s^{-1}}$, $\overline{\mu}_r = 24.31$-26.33 $\mathrm{mag~arcsec^{-2}}$, and $M_{*} = 10^{10.7-12.2}~M\mathrm{_{\odot}}$. The samples comprise 14 CGs identified at redshift $z=0.06$, 15 CGs at $z=0.14$ and 23 CGs at $z=0.23$. All of these systems constitute the $\sim$10\% of most compact groups except the sample of $z=0.23$, which represents the $\sim$15\% tail of the compactness distribution.

\subsection{Evolution of compact groups} \label{sec:evo_CG}

\begin{figure*}
    \centering
    \includegraphics[width=\textwidth]{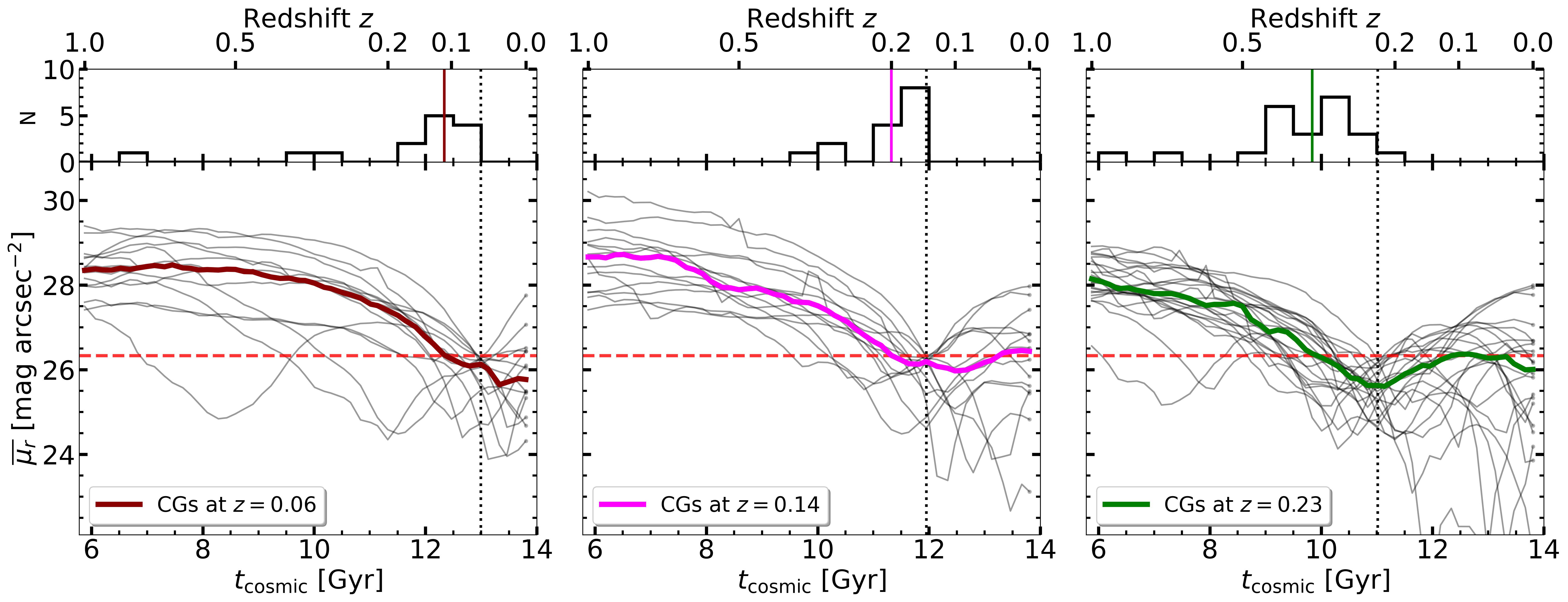}
    \caption{Evolution of compact group mean surface brightness, split by CG identification redshift. Histograms of threshold times are shown in the top panels. From left to right we show CGs identified at redshift $z=0.06$, $z=0.14$ and $z=0.23$. The median of each is shown as a coloured solid line (dark red, magenta and green, respectively). The vertical dotted black line indicates the identification time of each sample. CGs at different epochs evolve similarly: most cross the surface brightness threshold less than 1 Gyr before identification, reach maximum compactness shortly after, and then galaxies begin to orbit within the group. Post-identification evolution is highly diverse, reflecting the chaotic nature of these systems.}
\label{fig:median_z-geq-0}
\end{figure*}

We extend the analysis of the assembly of CGs to study the temporal evolution of CGs identified at $z>0$. In Fig. \ref{fig:median_z-geq-0} we show the median compactness evolution and threshold times for the three samples selected. At early times ($6 \lesssim t_{\mathrm{cosmic}}/\mathrm{Gyr} \lesssim 8$), most systems exhibit low surface brightness. After a visual inspection of the members' trajectories, we found that galaxies at these epochs were in looser configurations or not even in the same group. This relatively late assembly, combined with typical threshold times indicating that most of these groups become compact only about $\sim$$1$ Gyr before identification (consistent with the results for CGs$_{z=0}$, Sect. \ref{sec:assembly_CG}), shows that CGs originate from the sequential infall of galaxies into the same halo, where they gradually get closer until they reach a maximum compactness configuration. This reinforces the arbitrariness of the threshold value $\overline{\mu}_r = 26.33 ~ \mathrm{mag~arcsec^{-2}}$, as the maximum surface brightness of each CG depends on the impact parameter of the infalling galaxies.

Since the identification occurs at redshifts $z > 0$, we now study how already assembled CGs evolve until $z=0$. There is no unique pathway for the evolution of CGs, where some groups become looser systems again, and a few reach extremely high surface brightness values. While the former cases corresponds to transient CGs, which galaxies are in high compactness configurations only for a brief period of time, the latter cases correspond to systems that end with less than four galaxies, where mergers within members occur after the first close encounter, and they begin to orbit each other. In total, the median of each sample suggest that the majority of CGs meet the compactness criterion for about $\sim$2 Gyr.

\begin{figure}
    \centering
    \includegraphics[width=\columnwidth]{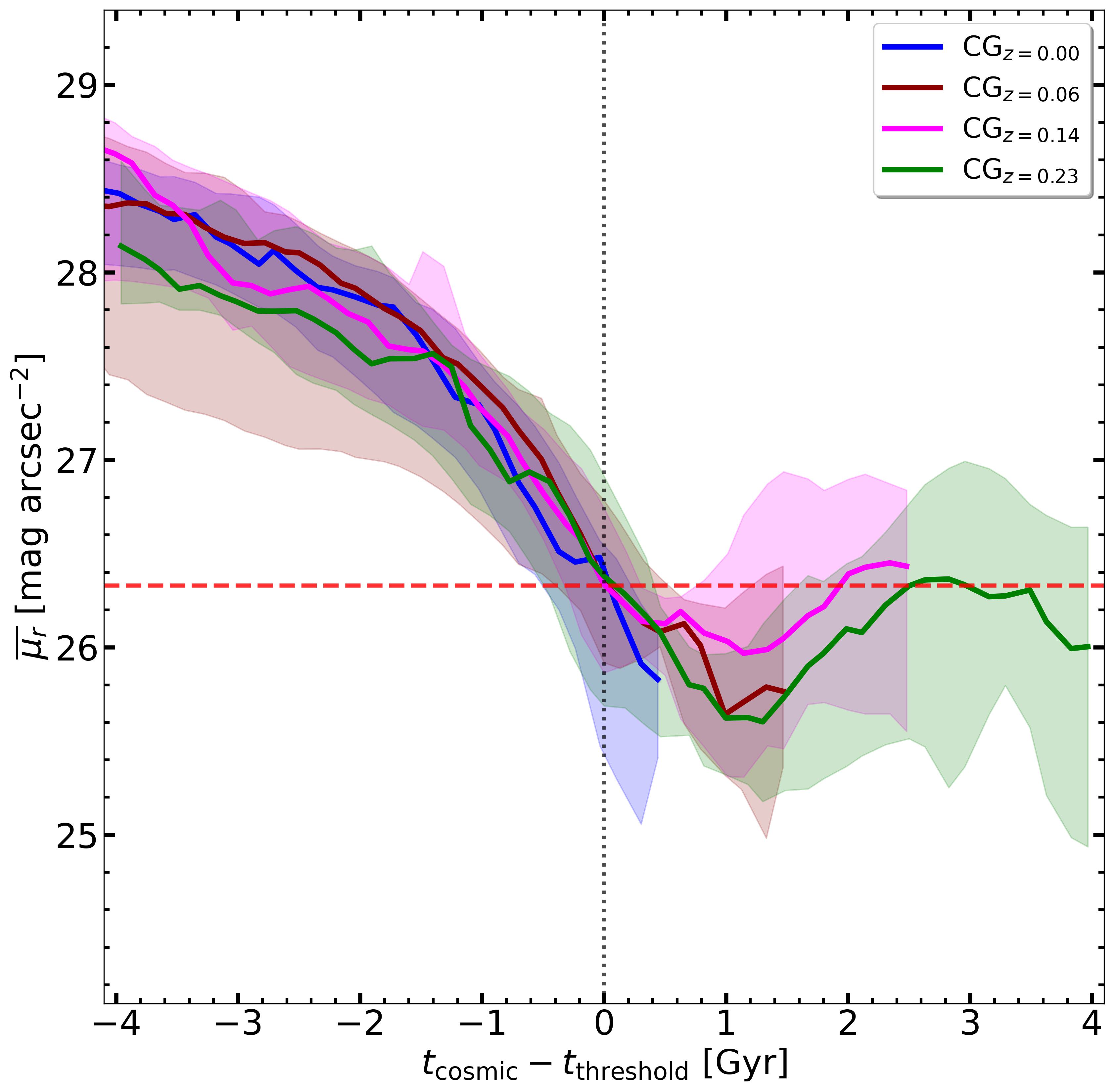}
    \caption{Evolution of compact group mean surface brightness in time relative to median threshold crossing. Shaded regions encompasses the 25th-75th percentiles. Colours indicate samples identified at redshift $z=0.00$ (blue), $z=0.06$ (dark red), $z=0.14$ (magenta) and $z=0.23$ (green). The last point of each curve corresponds to the final snapshot, at $t_{\mathrm{cosmic}} \approx 13.8$ Gyr. The red dashed horizontal line shows the threshold $\overline{\mu}_r = 26.33 ~ \mathrm{mag~arcsec^{-2}}$ and the black dotted vertical line indicates $t_{\mathrm{cosmic}} = t_{\mathrm{threshold}}$. When shifted horizontally, all medians follow a similar trend: Galaxies begin to inhabit the same halo, leading to the group assembly, a period of extreme compactness and subsequent orbital evolution.}
\label{fig:median_shifted}
\end{figure}

Since the cosmic time of identification only affects sample selection, in Fig. \ref{fig:median_shifted} we show the median curves for all analysed samples in this work ($z=0.00, ~0.06, ~0.14$ \&$~0.23$), shifted by their median threshold time, i.e. when 50\% of groups from a sample becomes compact. There is a clear similarity between the median of different samples, with most groups showing surface brightnesses as low as $\overline{\mu}_r \gtrsim 28 ~ \mathrm{mag~arcsec^{-2}}$ at $t_{\mathrm{cosmic}} - t_{\mathrm{thr}} \lesssim - 3$ Gyr. This corresponds to CGs not having all their members inhabiting the same halo until 1-2 Gyr before compaction, as previously mentioned. At those times, the scatter is almost constantly $\Delta \overline{\mu}_r \approx 1 ~ \mathrm{mag~arcsec^{-2}}$, until groups reach a maximum compactness of 25.5-26.0 $\mathrm{mag~arcsec^{-2}}$. Typically, this is achieved at $t_{\mathrm{cosmic}} - t_{\mathrm{thr}} \approx 1.5$ Gyr, which is arbitrary due to the fixed threshold value and, thus, it affects the time elapsed from identification to maximum group surface brightness. CGs$_{z=0}$ (blue solid line) will follow the same trend, but they have not evolved enough to reach maximum compactnesses. All medians stop at $t_{\mathrm{cosmic}} \approx 13.8$ Gyr, since it is the last snapshot of the simulation.

After the maximum surface brightness is achieved, the scatter within samples increases significantly ($\Delta \overline{\mu}_r \gtrsim 1.5 ~ \mathrm{mag~arcsec^{-2}}$) due to the diversity of evolutionary pathways of these extremely compact systems. CGs identified at $t_{\mathrm{cosmic}} \lesssim 12$ Gyr (CGs$_{z=0.14}$ and CGs$_{z=0.23}$), have enough time to evolve and spuriously satisfy multiple times the compactness criterion through time due to orbital motions of galaxies within groups.

In Sect. \ref{sec:CG_z0} we demonstrate that CGs are not a distinct population of galaxy groups, but rather the smallest groups at a given total stellar mass. Their surface brightness evolution reveals them to be transient configurations of galaxies. Median curves indicate that most are undergoing their first pericenter passage, resulting in a temporary peak in surface brightness shortly after compaction. This is supported by the fact that most CGs were in looser configurations 1-2 Gyr prior to identification, with only a few having early threshold times and experiencing a second phase of high surface brightness. Most groups of four or more similar-mass galaxies are recently assembled and cannot maintain a compact configuration for several Gyr due to orbital motion and/or mergers between members.

\subsection{Survival of compact groups by $z=0$} \label{sec:hist_crits}

While Fig. \ref{fig:median_z-geq-0} indicate that most CGs$_{z>0}$ end up in as configurations of galaxies sufficiently dense to fulfil the compactness criterion, fewer CGs are finally identified at $z=0$. Furthermore, the threshold times also indicate that CGs form statistically shortly before identification. To understand this deficit of long-lived CGs, we must consider not only surface brightness evolution but also member interactions and accretion of galaxies into these isolated systems. Using the identification criteria from Sect. \ref{sec:criteria} (richness, compactness and isolation), we consider both the occurrence of mergers between members and the number of galaxies with similar stellar mass that inhabits the same halo as the CG after identification, to elucidate what causes a group of galaxies to stop being a CG. For instance, if the central of a CG is no longer the central of the host halo after a few Gyr, the system can no longer be considered as isolated (e.g. due to the infall into a cluster).

\begin{figure}
    \centering
    \includegraphics[width=\columnwidth]{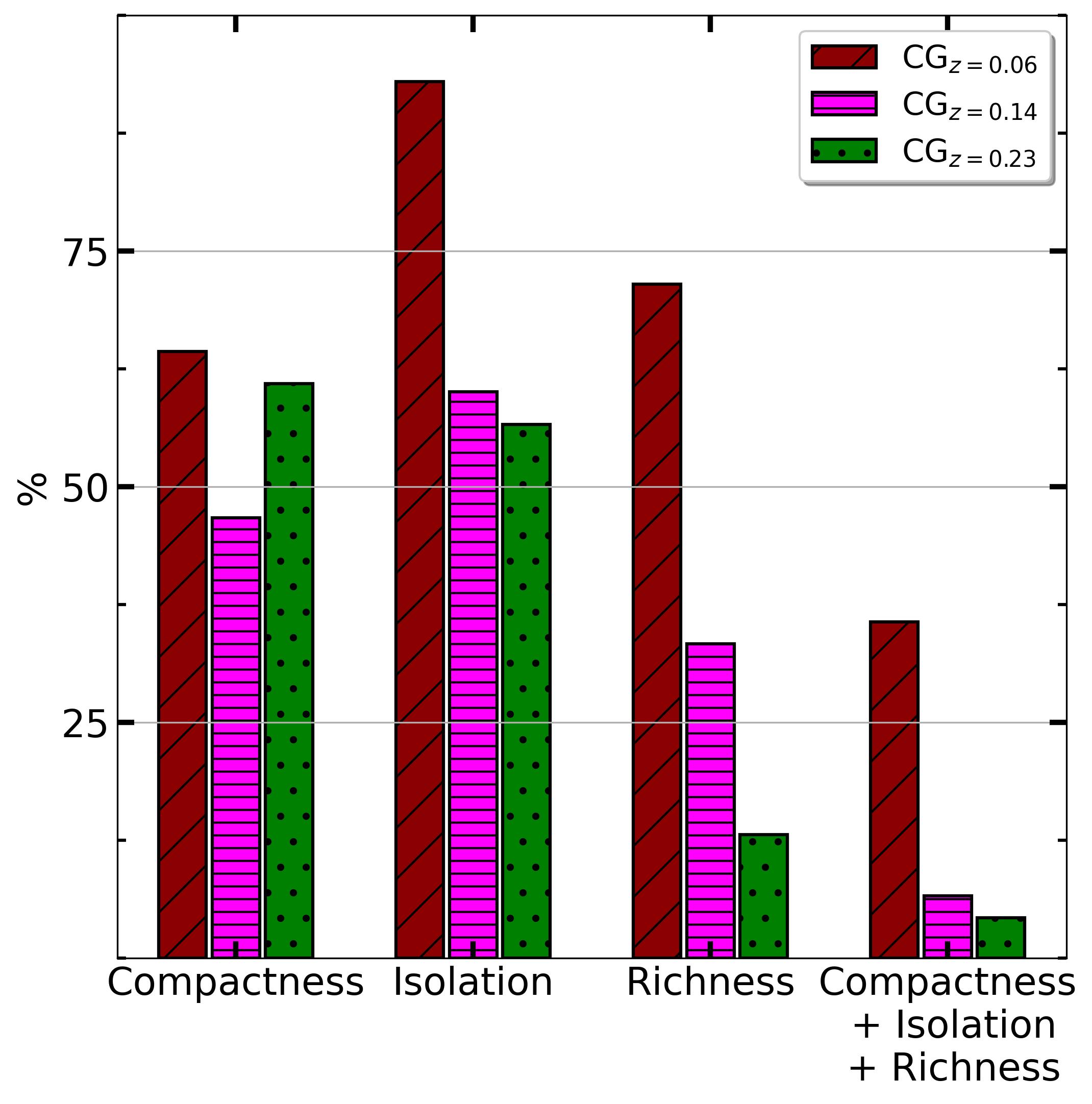}
    \caption{Percentages of compact groups meeting a given CG selection criterion at $z=0$, split by CG identification epoch. The sample of $z=0.06$ is shown in dark red, the one of $z=0.14$ in magenta and that of $z=0.23$ in green. The criteria analysed is compactness, isolation, richness and all three simultaneously, thus being identified as CGs at $z=0$. While 68\% of CGs$_{z>0}$ remain isolated by $z=0$, only 33\% are still conformed by at least four similar-mass galaxies. A small fraction of CGs remain as such at $z=0$, with lower values shown by older samples.}
\label{fig:hist_crits}
\end{figure}

We study the occurrence of mergers and group isolation by redshift $z=0$, tracing the \texttt{SUBLINK}'s main descendants of each galaxy of CGs$_{z>0}$. We then determine how many CGs remain as isolated groups of four or more similar-mass galaxies in a compact configuration (i.e. fully satisfying all criteria for the inclusion in the CGs$_{z=0}$ sample), measuring the fraction of groups that satisfy each criterion to identify the primary driver of the CGs abundances measured in TNG100. In Fig. \ref{fig:hist_crits} we show the percentage of each sample (colour-coded by identification time, as in Fig. \ref{fig:median_z-geq-0}) that satisfies each criterion at $z=0$ and all three simultaneously. Each sample is analysed separately due to differing evolution times. For example, some CGs$_{z=0.23}$ had enough time ($\Delta t \approx 3$ Gyr) to experience a second high-compactness phase\footnote{This also can be inferred from the median surface brightness evolution in Fig. \ref{fig:median_z-geq-0}.} and thus exhibit higher fraction of groups that satisfy the compactness criterion than CGs$_{z=0.14}$.

Considering the two oldest samples (CGs$_{z=0.14}$ in magenta hatched bars, and CGs$_{z=0.23}$ in green dotted bars), their least fulfilled criterion is the richness one, suggesting that this is the main driver of the low percentage of long-lived CGs remaining by $z=0$. Only 1/23 (4\%) of CGs$_{z=0.23}$ and 1/15 (7\%) of CGs$_{z=0.14}$ persist as CGs. Thus, most systems do not remain as CGs long after identification due to mergers between members. However, although the majority\footnote{67\% of CGs$_{z=0.14}$ and 87\% of CGs$_{z=0.23}$} of CGs identified at $z>0.1$ fail the richness criterion at $z=0$, strikingly only one becomes a single, isolated massive galaxy (we present more details in Sect. \ref{sec:coalescence}). Finally, 68\% of CGs$_{z>0}$ remain isolated by $z=0$, i.e. only 32\% fail the isolation criterion, experiencing the accretion of at least one galaxy with similar or greater stellar mass than the CG's central that ends up inhabiting the same halo. No CG undergoes a "split" or "fragmentation", where members end in different host haloes, as found in observed CGs \citep[cf.][]{Zheng2021, Tricottet2025}.

In a $\Lambda$CDM Universe, where structure formation proceeds hierarchically, accretion of massive galaxies into groups is expected within a Hubble time \citep[see e.g.][]{Genel2010,R-G2015}. Our results suggest that, after group assembly and compaction, mergers between members occur more rapidly than the accretion of galaxies with similar or greater mass than the central's. This is supported by the higher fraction of CGs failing the richness criterion rather than the isolation criterion by $z=0$. Nevertheless, mergers remain relatively infrequent, with most systems persisting as triplets after the first merger. This may be attributed to the relatively high velocity dispersions at high masses \citep[see e.g.][]{KontorovichKrivitsky1997,MakinoHut1997}.

The frequency of mergers between members within an already assembled CG is thus not trivial and is thoroughly explored in the next subsection.

\subsection{Merger rate and coalescence of compact groups}
\label{sec:coalescence}

\begin{figure}
    \centering
    \includegraphics[width=0.9\columnwidth]{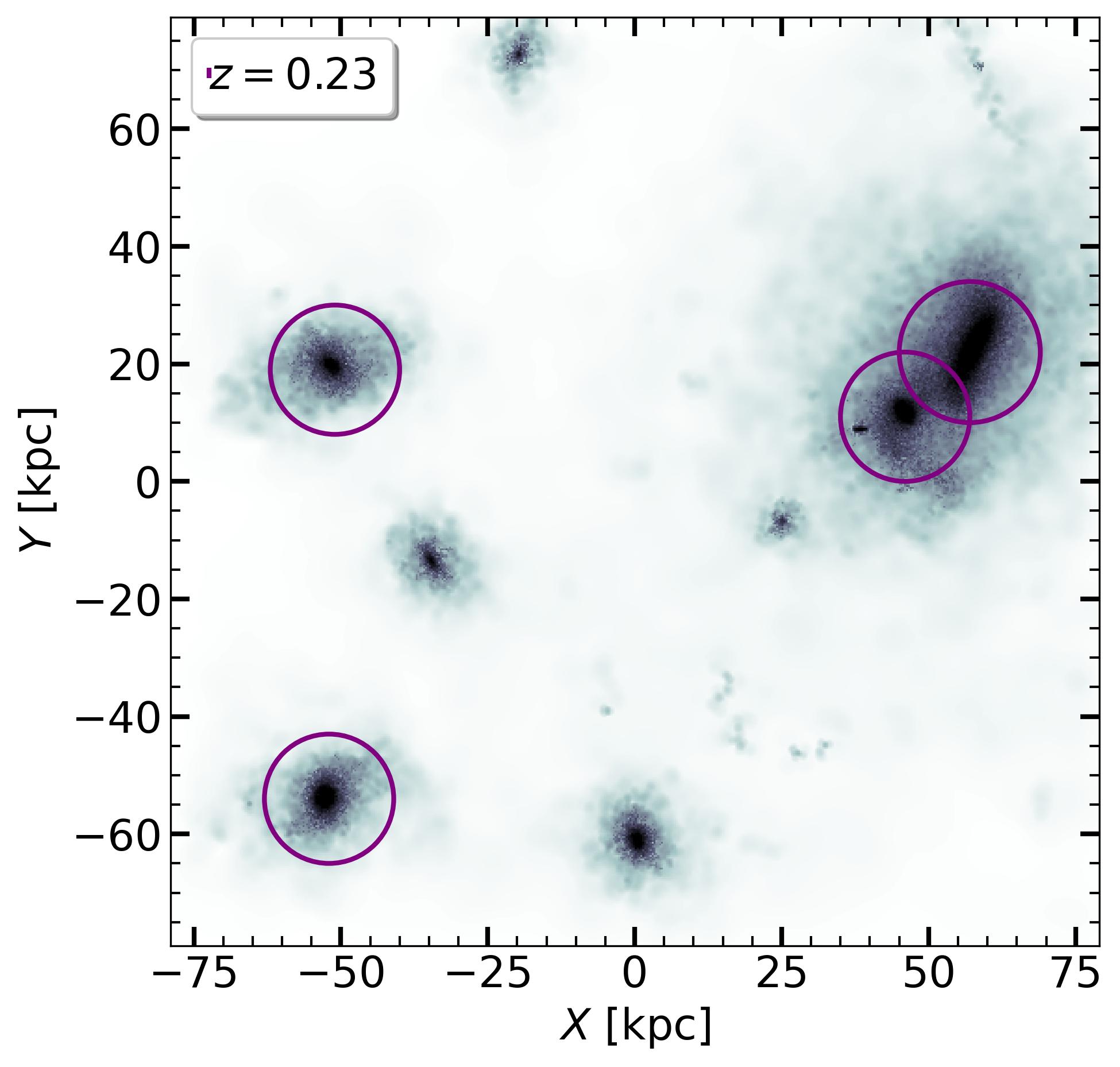}
    \includegraphics[width=0.9\columnwidth]{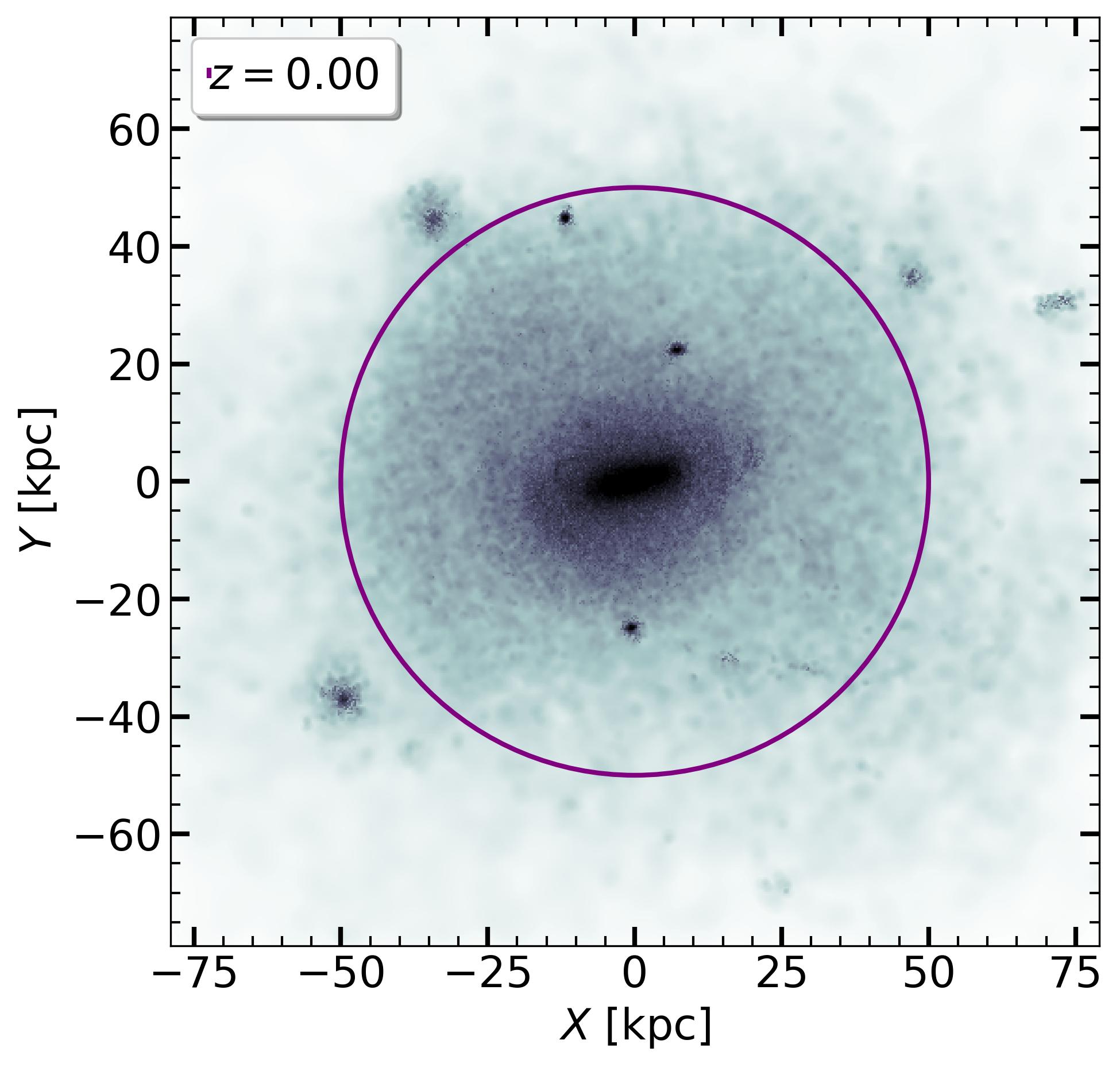}
    
    \caption{Two snapshots of the coalescing compact group: when identified (top) and at present (bottom). This is the only group studied which all members merge into a single galaxy before $z=0$. The purple circles indicate the four members of the CG, while other dark regions indicate dwarf galaxies that do not have stellar mass similar to the members of the CG. This CG$_{z=0.23}$ is the result of a close encounter between two pairs of galaxies: the two at the left ($X \approx -50$ kpc), and the two at the right ($X \approx 60$ kpc), which are very close in projection and show some signs of tidal interaction. Darker colours indicate a higher number density of particles. Images generated with the Py-SPH Viewer library \citep{BenitezLlambay2017}.}
    \label{fig:sph_FG}
\end{figure}

The expected short crossing times of CGs indicates that their galaxies should merge altogether in a few Gyr. However, in TNG100 we found only 1 CG (identified at $z=0.23$) whose galaxies all merge, resulting in a single galaxy of stellar mass $M_{*} \approx 10^{11} ~ M\mathrm{_{\odot}}$ without any satellites of $M_{*} > 10^{9} ~ M\mathrm{_{\odot}}$, being thus extremely isolated in its host halo. In Fig. \ref{fig:sph_FG} we show this CG at the identification time $t_{\mathrm{cosmic}} \approx 11.0$ Gyr ($z=0.23$, top panel) and its remnant at $t_{\mathrm{cosmic}} \approx 13.8$ Gyr ($z=0$, bottom panel). The cyan circles highlight the four members of the CG that merge into a single, massive, isolated galaxy at $z=0$. After a visual inspection of the trajectories of these four galaxy members, we found that these are two pairs of galaxies that begins to inhabit the same halo and approach each other head-on until they get close enough to be identified as a CG. Shortly after, each pair merge, resulting in two massive galaxies. Finally, these two galaxies end up merging together and the final galaxy remains in the halo without any non-dwarf satellite. The main characteristics and assembly history of this galaxy resembles to that attributed to Fossil Groups \citep{Ponman1994,Jones2003}, which harbour hot gas with the extent and other properties expected for groups, but the optical light is completely dominated by a single luminous, giant elliptical galaxy \citep[for a comprehensive review on the relation between fossil groups and compact groups see][]{Farhang2017, Aguerri2021}.

Analysing each of the 52 CGs identified at redshift $z>0$, one ends up as a single galaxy (shown in Fig. \ref{fig:sph_FG}), none as a pair of galaxies and the other 51 remain as systems of three or more galaxies by $z=0$. Within the latter, the majority remained isolated and a few became substructures of larger systems. Since the only CG to coalesce belongs to the CG$_{z=0.23}$ sample, we further study the complete population of galaxy groups from $z=0.23$ (including both compacts and non-compacts), measuring the frequency of mergers throughout their evolution. Although there are some groups that experience more than one merger, the CG shown in Fig. \ref{fig:sph_FG} is the only one\footnote{Labelled as FoF group \#345 at redshift $z=0.23$ and FoF group \#341 at redshift $z=0$.} whose members coalesced. 

\begin{figure}
    \includegraphics[width=\columnwidth]{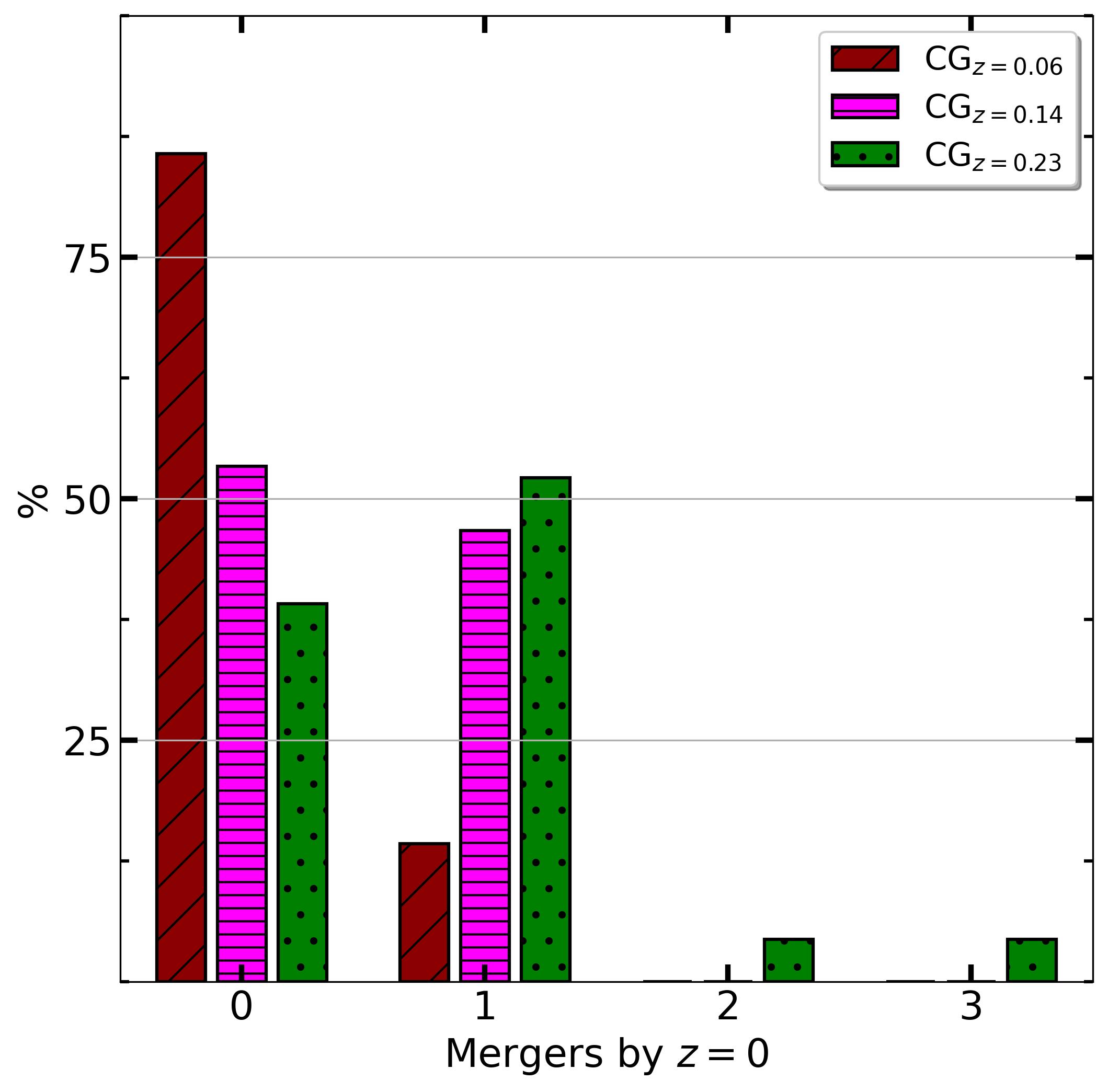}
    \caption{Distribution of number of galaxy mergers within compact groups, split by identification epoch. While only 2/15 ($\sim$14\%) of CGs$_{z=0.06}$ experience one merger, the CGs$_{z=0.23}$ sample is the only where most systems had experienced at least one merger. Coalescence of CGs that had $\Delta t < 3$ Gyr to evolve is somewhat infrequent.}
    \label{fig:merger_rate}
\end{figure}

Taking into account all mergers between galaxy members from CGs$_{z>0}$, we found a rate of $(0.24 \pm 0.32)$ mergers/Gyr in these systems\footnote{These values are the average and standard deviation.}. As shown in Fig. \ref{fig:merger_rate} (colour-coded by sample, as in Fig. \ref{fig:hist_crits}), the CGs$_{z=0.23}$ sample is the only one where the majority of groups experienced at least one merger (14/23, 61\%). For those CGs$_{z>0}$ that experienced at least one merger (23/52, 44\%), the first one occurs $(0.92 \pm 0.46)$ Gyr after identification and it is the only one in $\sim$91\% of cases. The apparent extreme rarity of CGs (and groups in general) that accomplish coalescence suggest that the time needed to merge four or more massive galaxies in a cosmological environment is $\Delta t \gtrsim 3$ Gyr (cosmic time between $z=0.23$ and $z=0$). Further studies of the initial conditions and properties of these particular cases of groups that fully coalesce is needed and it is out of the scope of this paper.

\section{Discussion} \label{sec:discussion}

The criteria described in Sec. \ref{sec:criteria} allow us to study physically compact systems constituted by 4 or more galaxies with similar stellar mass, including the central of the host halo. These criteria avoid contamination from spurious groups that can appear compact in projection, and configurations of galaxies that are substructures within larger systems and thus not isolated. We limit the definition of "similar mass" to one order of magnitude difference (1 dex) in stellar mass rather than the commonly used 3 magnitudes (1.2 dex in luminosity), aiming for the analysis of the survival of systems with members with mass within an order of magnitude. Furthermore, tidal interactions induce star formation and thus enhance galaxy luminosities, which relies on the photometric band choice and the assumed stellar population synthesis models of the simulation.

Given these caveats, the properties of our CG samples are fairly comparable with both theoretical and observational work. First, CGs are not a special sub-type of groups, since compared to all isolated and rich groups, they are only the $\sim$10\% tail of the distribution of group compactness, being the smallest at given stellar mass. There is no segregation on both the compactness and velocity dispersion between CGs and non-compact groups, in agreement with the observational results shown by e.g. \citet{Zheng2022,Tricottet2025}.

The 3D compactness criterion ensures galaxies in physically compact configurations when averaged over 3 orthogonal projections, using as projected size the maximum distance of a member to the geometric centre of the system. This radius is a factor 1-1.4 larger than the smallest circumscribed circle, which is commonly used to define the size of a system. However, the main properties of the CGs identified are robust to different definitions of group size.

Limiting the search to a single axis and measuring the fraction of truly CGs within all the identified, we found that 60-70\% of CG in projection are truly compact. This fraction is larger than the one found by \citet{DG2020} using semi-analytical models on gravity-only simulations and by \citet{HyP2020} using the EAGLE cosmological hydrodynamical simulation. The key differences may reside on the richness and similarity criteria used in these works, where systems with fewer galaxies ($N \geq 3$) and a brightness difference of $\Delta m \leq 3$ mag are allowed. In addition, the previously mentioned authors performed their analysis in redshift space, which is more prone to projection effects in comparison with searches in real space, as we used in this work.

Lastly, we studied the temporal evolution within $1 \geq z \geq 0$ of the simulated CGs. Most systems start to meet the selection criteria about 1 Gyr before identification, and only a few are going through a second phase of CG. Regarding formation channels, \citet[][]{DG2021} analysed lightcones derived from semi-analytical models previously run on gravity-only simulations and
found that less than 10\% of true CG formed at early times and remained as such over several Gyr. After a visual inspection of the trajectories of the members, we found that the main mechanisms of CG formation are: i) when a fourth galaxy enters the same host halo of an already formed triplet of galaxies; and ii) when two pairs of galaxies are having a close encounter, thus satisfying the compactness criterion at the identification time.

Following these systems forward in time, we found that most are CG only about 1-3 Gyr in total. Shortly after group assembly, members can merge or infall into other massive structures. Only one group remained rich $(N \geq 4)$, isolated and compact after 3 Gyr of cosmic evolution, and also only one fully accomplished coalescence (all members merge altogether). This suggests that i) survivability of CGs is lower than the predicted by \citet{DG2021}; and ii) the coalescence time of CGs is longer than the results shown by \citet{HyP2020}. These inconsistencies with other theoretical works may be attributed to i) use similar mass rather than similar luminosity, since luminosity could be a bad tracer of mass when bursts of star formation are triggered by tidal interactions, and galaxies with similar mass are more affected by dynamical friction within the same host halo thus a higher frequency of mergers is expected; and ii) the richness criterion, because mergers can happen more frequently, but our systems consist of at least 4 massive galaxies that may need more time to fully coalesce compared to the CGs of 3 galaxies that dominate the sample of \citet{HyP2020} in the EAGLE simulation.

\section{Summary and Conclusions} \label{sec:conclusion}

In this work, we used the TNG100 hydrodynamic cosmological simulation to characterize the evolution of compacts groups of galaxies identified at low redshifts ($0 \leq z \leq 0.23$). We defined compact groups as systems with four or more similar-mass galaxies inhabiting the same halo and exhibiting high group surface brightness in three orthogonal projections. Our derived abundance of these systems is consistent with the space density of compact groups identified in the SDSS-DR12 \citep{Sohn2015}.

We analysed the surface brightness evolution of four compact groups samples across cosmic times $5.9 \lesssim t\mathrm{_{cosmic}/Gyr} \lesssim 13.8$, measuring their typical compaction timescale. We found that most compact groups achieve high compactness less than 1 Gyr before identification, while others assembled several Gyrs ago and are experiencing a second phase of compaction. Statistically, all compact groups follow a similar evolutionary path: galaxies begin to inhabit the same halo (group assembly), reach a maximum group surface brightness (associated with a first close encounter), and then begin orbiting one another.

Tracking the forward temporal evolution of compact groups identified at $z>0$, we found that 89\% fail at least one identification criterion before $z=0$. Mergers are the primary driver of this disruption, particularly in groups evolving for more than 2 Gyr. This indicates a higher merger rate between group members than the accretion rate of new massive galaxies into the group halo. Notably, we observed only one group that achieved complete coalescence, where all galaxies progressively merged into a single, massive, and extremely isolated galaxy.

Our analysis suggest that compact groups are transient configurations of groups of galaxies, unable to remain as isolated systems of four or more similar-mass galaxies in a physically compact configuration, and thus satisfying the identification criteria only for short periods. However, the high frequency of mergers compared to the accretion of new massive galaxies makes compact groups excellent laboratories for studying the effects of galaxy tidal interactions within a $\Lambda$CDM Universe. Given the brief duration of the compact group phase, their abundance throughout cosmic evolution can be explained by the close encounters of several similar-mass galaxies, representing one more step in the hierarchical formation of cosmic structures.

\begin{acknowledgements}
    We thank the referee Gary Mamon for an excellent and a very constructive report of the manuscript. The authors are also thankful to Julio Navarro for helpful discussions that helped shape this paper. This work was partially supported by the Consejo de Investigaciones Científicas y Técnicas de la República Argentina (CONICET) and the Secretaría de Ciencia y Técnica de la Universidad Nacional de Córdoba (SeCyT). The IllustrisTNG simulations were undertaken with compute time awarded by the Gauss Centre for Supercomputing (GCS) under GCS Large-Scale Projects GCS-ILLU and GCS-DWAR on the GCS share of the supercomputer Hazel Hen at the High Performance Computing Center Stuttgart (HLRS), as well as on the machines of the Max Planck Computing and Data Facility (MPCDF) in Garching, Germany. JAB is grateful for partial financial support from NSF-CAREER-1945310 and NSF-AST-2107993 grants.
\end{acknowledgements}

\bibliographystyle{aa}
\bibliography{aa55375-25}

\begin{thebibliography}{65}
\expandafter\ifx\csname natexlab\endcsname\relax\def\natexlab#1{#1}\fi

\bibitem[{{Aguerri} \& {Zarattini}(2021)}]{Aguerri2021}
{Aguerri}, J. A.~L. \& {Zarattini}, S. 2021, Universe, 7, 132

\bibitem[{{Anbajagane} {et~al.}(2022){Anbajagane}, {Aung}, {Evrard}, {Farahi}, {Nagai}, {Barnes}, {Cui}, {Dolag}, {McCarthy}, {Rasia}, \& {Yepes}}]{Anbajagane2022}
{Anbajagane}, D., {Aung}, H., {Evrard}, A.~E., {et~al.} 2022, \mnras, 510, 2980

\bibitem[{{Athanassoula} {et~al.}(1997){Athanassoula}, {Makino}, \& {Bosma}}]{Athanassoula1997}
{Athanassoula}, E., {Makino}, J., \& {Bosma}, A. 1997, \mnras, 286, 825

\bibitem[{{Barnes}(1989)}]{Barnes1989}
{Barnes}, J.~E. 1989, \nat, 338, 123

\bibitem[{{Ben{\'\i}tez-Llambay}(2017)}]{BenitezLlambay2017}
{Ben{\'\i}tez-Llambay}, A. 2017, {Py-SPHViewer: Cosmological simulations using Smoothed Particle Hydrodynamics}, Astrophysics Source Code Library, record ascl:1712.003

\bibitem[{{Carnevali} {et~al.}(1981){Carnevali}, {Cavaliere}, \& {Santangelo}}]{Carnevali1981}
{Carnevali}, P., {Cavaliere}, A., \& {Santangelo}, P. 1981, \apj, 249, 449

\bibitem[{{Chabrier}(2003)}]{Chabrier2003}
{Chabrier}, G. 2003, \pasp, 115, 763

\bibitem[{{Davis} {et~al.}(1985){Davis}, {Efstathiou}, {Frenk}, \& {White}}]{Davies1985}
{Davis}, M., {Efstathiou}, G., {Frenk}, C.~S., \& {White}, S.~D.~M. 1985, \apj, 292, 371

\bibitem[{{Diaferio} {et~al.}(1994){Diaferio}, {Geller}, \& {Ramella}}]{Diaf1994}
{Diaferio}, A., {Geller}, M.~J., \& {Ramella}, M. 1994, \aj, 107, 868

\bibitem[{{D{\'\i}az-Gim{\'e}nez} \& {Mamon}(2010)}]{DG2010}
{D{\'\i}az-Gim{\'e}nez}, E. \& {Mamon}, G.~A. 2010, \mnras, 409, 1227

\bibitem[{{D{\'\i}az-Gim{\'e}nez} {et~al.}(2012){D{\'\i}az-Gim{\'e}nez}, {Mamon}, {Pacheco}, {Mendes de Oliveira}, \& {Alonso}}]{DG2012}
{D{\'\i}az-Gim{\'e}nez}, E., {Mamon}, G.~A., {Pacheco}, M., {Mendes de Oliveira}, C., \& {Alonso}, M.~V. 2012, \mnras, 426, 296

\bibitem[{{D{\'\i}az-Gim{\'e}nez} {et~al.}(2020){D{\'\i}az-Gim{\'e}nez}, {Taverna}, {Zandivarez}, \& {Mamon}}]{DG2020}
{D{\'\i}az-Gim{\'e}nez}, E., {Taverna}, A., {Zandivarez}, A., \& {Mamon}, G.~A. 2020, \mnras, 492, 2588

\bibitem[{{D{\'\i}az-Gim{\'e}nez} {et~al.}(2021){D{\'\i}az-Gim{\'e}nez}, {Zandivarez}, \& {Mamon}}]{DG2021}
{D{\'\i}az-Gim{\'e}nez}, E., {Zandivarez}, A., \& {Mamon}, G.~A. 2021, \mnras, 503, 394

\bibitem[{{D{\'\i}az-Gim{\'e}nez} {et~al.}(2018){D{\'\i}az-Gim{\'e}nez}, {Zandivarez}, \& {Taverna}}]{DG2018}
{D{\'\i}az-Gim{\'e}nez}, E., {Zandivarez}, A., \& {Taverna}, A. 2018, \aap, 618, A157

\bibitem[{{Dolag} {et~al.}(2009){Dolag}, {Borgani}, {Murante}, \& {Springel}}]{Dolag2009}
{Dolag}, K., {Borgani}, S., {Murante}, G., \& {Springel}, V. 2009, \mnras, 399, 497

\bibitem[{{Farhang} {et~al.}(2017){Farhang}, {Khosroshahi}, {Mamon}, {Dariush}, \& {Raouf}}]{Farhang2017}
{Farhang}, A., {Khosroshahi}, H.~G., {Mamon}, G.~A., {Dariush}, A.~A., \& {Raouf}, M. 2017, \apj, 840, 58

\bibitem[{{Genel} {et~al.}(2010){Genel}, {Bouch{\'e}}, {Naab}, {Sternberg}, \& {Genzel}}]{Genel2010}
{Genel}, S., {Bouch{\'e}}, N., {Naab}, T., {Sternberg}, A., \& {Genzel}, R. 2010, \apj, 719, 229

\bibitem[{{Governato} {et~al.}(1991){Governato}, {Bhatia}, \& {Chincarini}}]{Governato1991}
{Governato}, F., {Bhatia}, R., \& {Chincarini}, G. 1991, \apjl, 371, L15

\bibitem[{{Hartsuiker} \& {Ploeckinger}(2020)}]{HyP2020}
{Hartsuiker}, L. \& {Ploeckinger}, S. 2020, \mnras, 491, L66

\bibitem[{{Hernquist} {et~al.}(1995){Hernquist}, {Katz}, \& {Weinberg}}]{Hernquist1995}
{Hernquist}, L., {Katz}, N., \& {Weinberg}, D.~H. 1995, \apj, 442, 57

\bibitem[{{Hickson} {et~al.}(1992){Hickson}, {Mendes de Oliveira}, {Huchra}, \& {Palumbo}}]{Hickson1992}
{Hickson}, P., {Mendes de Oliveira}, C., {Huchra}, J.~P., \& {Palumbo}, G.~G. 1992, \apj, 399, 353

\bibitem[{{Hickson} \& {Rood}(1988)}]{HicksonRood1988}
{Hickson}, P. \& {Rood}, H.~J. 1988, \apjl, 331, L69

\bibitem[{{Jones} {et~al.}(2003){Jones}, {Ponman}, {Horton}, {Babul}, {Ebeling}, \& {Burke}}]{Jones2003}
{Jones}, L.~R., {Ponman}, T.~J., {Horton}, A., {et~al.} 2003, \mnras, 343, 627

\bibitem[{{Krivitsky} \& {Kontorovich}(1997)}]{KontorovichKrivitsky1997}
{Krivitsky}, D.~S. \& {Kontorovich}, V.~M. 1997, \aap, 327, 921

\bibitem[{{Makino} \& {Hut}(1997)}]{MakinoHut1997}
{Makino}, J. \& {Hut}, P. 1997, \apj, 481, 83

\bibitem[{{Mamon}(1986)}]{Mamon1986}
{Mamon}, G.~A. 1986, \apj, 307, 426

\bibitem[{{Mamon}(1987)}]{Mamon1987}
{Mamon}, G.~A. 1987, \apj, 321, 622

\bibitem[{{Mamon}(1992)}]{Mamon1992}
{Mamon}, G.~A. 1992, \apjl, 401, L3

\bibitem[{{Mamon}(2008)}]{Mamon2008}
{Mamon}, G.~A. 2008, \aap, 486, 113

\bibitem[{{Mamon} {et~al.}(2013){Mamon}, {Biviano}, \& {Bou{\'e}}}]{Mamon2013}
{Mamon}, G.~A., {Biviano}, A., \& {Bou{\'e}}, G. 2013, \mnras, 429, 3079

\bibitem[{{Marinacci} {et~al.}(2018){Marinacci}, {Vogelsberger}, {Pakmor}, {Torrey}, {Springel}, {Hernquist}, {Nelson}, {Weinberger}, {Pillepich}, {Naiman}, \& {Genel}}]{Miranacci2018}
{Marinacci}, F., {Vogelsberger}, M., {Pakmor}, R., {et~al.} 2018, \mnras, 480, 5113

\bibitem[{{McConnachie} {et~al.}(2008){McConnachie}, {Ellison}, \& {Patton}}]{McC2008}
{McConnachie}, A.~W., {Ellison}, S.~L., \& {Patton}, D.~R. 2008, \mnras, 387, 1281

\bibitem[{{McConnachie} {et~al.}(2009){McConnachie}, {Patton}, {Ellison}, \& {Simard}}]{McC2009}
{McConnachie}, A.~W., {Patton}, D.~R., {Ellison}, S.~L., \& {Simard}, L. 2009, \mnras, 395, 255

\bibitem[{{Naiman} {et~al.}(2018){Naiman}, {Pillepich}, {Springel}, {Ramirez-Ruiz}, {Torrey}, {Vogelsberger}, {Pakmor}, {Nelson}, {Marinacci}, {Hernquist}, {Weinberger}, \& {Genel}}]{Naiman2018}
{Naiman}, J.~P., {Pillepich}, A., {Springel}, V., {et~al.} 2018, \mnras, 477, 1206

\bibitem[{{Nelson} {et~al.}(2019){Nelson}, {Springel}, {Pillepich}, {Rodriguez-Gomez}, {Torrey}, {Genel}, {Vogelsberger}, {Pakmor}, {Marinacci}, {Weinberger}, {Kelley}, {Lovell}, {Diemer}, \& {Hernquist}}]{Nelson2019}
{Nelson}, D., {Springel}, V., {Pillepich}, A., {et~al.} 2019, Computational Astrophysics and Cosmology, 6, 2

\bibitem[{{Pakmor} {et~al.}(2016){Pakmor}, {Springel}, {Bauer}, {Mocz}, {Munoz}, {Ohlmann}, {Schaal}, \& {Zhu}}]{Pakmor2016}
{Pakmor}, R., {Springel}, V., {Bauer}, A., {et~al.} 2016, \mnras, 455, 1134

\bibitem[{{Pillepich} {et~al.}(2018{\natexlab{a}}){Pillepich}, {Nelson}, {Hernquist}, {Springel}, {Pakmor}, {Torrey}, {Weinberger}, {Genel}, {Naiman}, {Marinacci}, \& {Vogelsberger}}]{Pill2018}
{Pillepich}, A., {Nelson}, D., {Hernquist}, L., {et~al.} 2018{\natexlab{a}}, \mnras, 475, 648

\bibitem[{{Pillepich} {et~al.}(2018{\natexlab{b}}){Pillepich}, {Nelson}, {Hernquist}, {Springel}, {Pakmor}, {Torrey}, {Weinberger}, {Genel}, {Naiman}, {Marinacci}, \& {Vogelsberger}}]{Pillepich2018}
{Pillepich}, A., {Nelson}, D., {Hernquist}, L., {et~al.} 2018{\natexlab{b}}, \mnras, 475, 648

\bibitem[{{Planck Collaboration} {et~al.}(2016){Planck Collaboration}, {Ade}, {Aghanim}, {Arnaud}, {Ashdown}, {Aumont}, {Baccigalupi}, {Banday}, {Barreiro}, {Bartlett}, {Bartolo}, {Battaner}, {Battye}, {Benabed}, {Beno{\^\i}t}, {Benoit-L{\'e}vy}, {Bernard}, {Bersanelli}, {Bielewicz}, {Bock}, {Bonaldi}, {Bonavera}, {Bond}, {Borrill}, {Bouchet}, {Boulanger}, {Bucher}, {Burigana}, {Butler}, {Calabrese}, {Cardoso}, {Catalano}, {Challinor}, {Chamballu}, {Chary}, {Chiang}, {Chluba}, {Christensen}, {Church}, {Clements}, {Colombi}, {Colombo}, {Combet}, {Coulais}, {Crill}, {Curto}, {Cuttaia}, {Danese}, {Davies}, {Davis}, {de Bernardis}, {de Rosa}, {de Zotti}, {Delabrouille}, {D{\'e}sert}, {Di Valentino}, {Dickinson}, {Diego}, {Dolag}, {Dole}, {Donzelli}, {Dor{\'e}}, {Douspis}, {Ducout}, {Dunkley}, {Dupac}, {Efstathiou}, {Elsner}, {En{\ss}lin}, {Eriksen}, {Farhang}, {Fergusson}, {Finelli}, {Forni}, {Frailis}, {Fraisse}, {Franceschi}, {Frejsel}, {Galeotta}, {Galli}, {Ganga}, {Gauthier}, {Gerbino}, {Ghosh}, {Giard},
  {Giraud-H{\'e}raud}, {Giusarma}, {Gjerl{\o}w}, {Gonz{\'a}lez-Nuevo}, {G{\'o}rski}, {Gratton}, {Gregorio}, {Gruppuso}, {Gudmundsson}, {Hamann}, {Hansen}, {Hanson}, {Harrison}, {Helou}, {Henrot-Versill{\'e}}, {Hern{\'a}ndez-Monteagudo}, {Herranz}, {Hildebrandt}, {Hivon}, {Hobson}, {Holmes}, {Hornstrup}, {Hovest}, {Huang}, {Huffenberger}, {Hurier}, {Jaffe}, {Jaffe}, {Jones}, {Juvela}, {Keih{\"a}nen}, {Keskitalo}, {Kisner}, {Kneissl}, {Knoche}, {Knox}, {Kunz}, {Kurki-Suonio}, {Lagache}, {L{\"a}hteenm{\"a}ki}, {Lamarre}, {Lasenby}, {Lattanzi}, {Lawrence}, {Leahy}, {Leonardi}, {Lesgourgues}, {Levrier}, {Lewis}, {Liguori}, {Lilje}, {Linden-V{\o}rnle}, {L{\'o}pez-Caniego}, {Lubin}, {Mac{\'\i}as-P{\'e}rez}, {Maggio}, {Maino}, {Mandolesi}, {Mangilli}, {Marchini}, {Maris}, {Martin}, {Martinelli}, {Mart{\'\i}nez-Gonz{\'a}lez}, {Masi}, {Matarrese}, {McGehee}, {Meinhold}, {Melchiorri}, {Melin}, {Mendes}, {Mennella}, {Migliaccio}, {Millea}, {Mitra}, {Miville-Desch{\^e}nes}, {Moneti}, {Montier}, {Morgante}, {Mortlock},
  {Moss}, {Munshi}, {Murphy}, {Naselsky}, {Nati}, {Natoli}, {Netterfield}, {N{\o}rgaard-Nielsen}, {Noviello}, {Novikov}, {Novikov}, {Oxborrow}, {Paci}, {Pagano}, {Pajot}, {Paladini}, {Paoletti}, {Partridge}, {Pasian}, {Patanchon}, {Pearson}, {Perdereau}, {Perotto}, {Perrotta}, {Pettorino}, {Piacentini}, {Piat}, {Pierpaoli}, {Pietrobon}, {Plaszczynski}, {Pointecouteau}, {Polenta}, {Popa}, {Pratt}, {Pr{\'e}zeau}, {Prunet}, {Puget}, {Rachen}, {Reach}, {Rebolo}, {Reinecke}, {Remazeilles}, {Renault}, {Renzi}, {Ristorcelli}, {Rocha}, {Rosset}, {Rossetti}, {Roudier}, {Rouill{\'e} d'Orfeuil}, {Rowan-Robinson}, {Rubi{\~n}o-Mart{\'\i}n}, {Rusholme}, {Said}, {Salvatelli}, {Salvati}, {Sandri}, {Santos}, {Savelainen}, {Savini}, {Scott}, {Seiffert}, {Serra}, {Shellard}, {Spencer}, {Spinelli}, {Stolyarov}, {Stompor}, {Sudiwala}, {Sunyaev}, {Sutton}, {Suur-Uski}, {Sygnet}, {Tauber}, {Terenzi}, {Toffolatti}, {Tomasi}, {Tristram}, {Trombetti}, {Tucci}, {Tuovinen}, {T{\"u}rler}, {Umana}, {Valenziano}, {Valiviita}, {Van Tent},
  {Vielva}, {Villa}, {Wade}, {Wandelt}, {Wehus}, {White}, {White}, {Wilkinson}, {Yvon}, {Zacchei}, \& {Zonca}}]{Planck2016}
{Planck Collaboration}, {Ade}, P.~A.~R., {Aghanim}, N., {et~al.} 2016, \aap, 594, A13

\bibitem[{{Pompei} \& {Iovino}(2012)}]{PyI2012}
{Pompei}, E. \& {Iovino}, A. 2012, \aap, 539, A106

\bibitem[{{Ponman} {et~al.}(1994){Ponman}, {Allan}, {Jones}, {Merrifield}, {McHardy}, {Lehto}, \& {Luppino}}]{Ponman1994}
{Ponman}, T.~J., {Allan}, D.~J., {Jones}, L.~R., {et~al.} 1994, \nat, 369, 462

\bibitem[{{Rodriguez-Gomez} {et~al.}(2015){Rodriguez-Gomez}, {Genel}, {Vogelsberger}, {Sijacki}, {Pillepich}, {Sales}, {Torrey}, {Snyder}, {Nelson}, {Springel}, {Ma}, \& {Hernquist}}]{R-G2015}
{Rodriguez-Gomez}, V., {Genel}, S., {Vogelsberger}, M., {et~al.} 2015, \mnras, 449, 49

\bibitem[{{Rose}(1977)}]{Rose1977}
{Rose}, J.~A. 1977, \apj, 211, 311

\bibitem[{{Rose}(1979)}]{Rose1979}
{Rose}, J.~A. 1979, \apj, 231, 10

\bibitem[{{Schaye} {et~al.}(2015){Schaye}, {Crain}, {Bower}, {Furlong}, {Schaller}, {Theuns}, {Dalla Vecchia}, {Frenk}, {McCarthy}, {Helly}, {Jenkins}, {Rosas-Guevara}, {White}, {Baes}, {Booth}, {Camps}, {Navarro}, {Qu}, {Rahmati}, {Sawala}, {Thomas}, \& {Trayford}}]{EAGLE2015}
{Schaye}, J., {Crain}, R.~A., {Bower}, R.~G., {et~al.} 2015, \mnras, 446, 521

\bibitem[{{Schneider} \& {Gunn}(1982)}]{Schneider1982}
{Schneider}, D.~P. \& {Gunn}, J.~E. 1982, \apj, 263, 14

\bibitem[{{Sohn} {et~al.}(2015){Sohn}, {Hwang}, {Geller}, {Diaferio}, {Rines}, {Lee}, \& {Lee}}]{Sohn2015}
{Sohn}, J., {Hwang}, H.~S., {Geller}, M.~J., {et~al.} 2015, Journal of Korean Astronomical Society, 48, 381

\bibitem[{{Springel}(2010)}]{Springel2010}
{Springel}, V. 2010, \mnras, 401, 791

\bibitem[{{Springel} {et~al.}(2018){Springel}, {Pakmor}, {Pillepich}, {Weinberger}, {Nelson}, {Hernquist}, {Vogelsberger}, {Genel}, {Torrey}, {Marinacci}, \& {Naiman}}]{Springel2018}
{Springel}, V., {Pakmor}, R., {Pillepich}, A., {et~al.} 2018, \mnras, 475, 676

\bibitem[{{Springel} {et~al.}(2005){Springel}, {White}, {Jenkins}, {Frenk}, {Yoshida}, {Gao}, {Navarro}, {Thacker}, {Croton}, {Helly}, {Peacock}, {Cole}, {Thomas}, {Couchman}, {Evrard}, {Colberg}, \& {Pearce}}]{Springel2005}
{Springel}, V., {White}, S. D.~M., {Jenkins}, A., {et~al.} 2005, \nat, 435, 629

\bibitem[{{Springel} {et~al.}(2001){Springel}, {White}, {Tormen}, \& {Kauffmann}}]{Springel2001}
{Springel}, V., {White}, S. D.~M., {Tormen}, G., \& {Kauffmann}, G. 2001, \mnras, 328, 726

\bibitem[{{Taverna} {et~al.}(2016){Taverna}, {D{\'\i}az-Gim{\'e}nez}, {Zandivarez}, {Joray}, \& {Kanagusuku}}]{Taverna2016}
{Taverna}, A., {D{\'\i}az-Gim{\'e}nez}, E., {Zandivarez}, A., {Joray}, F., \& {Kanagusuku}, M.~J. 2016, \mnras, 461, 1539

\bibitem[{{Taverna} {et~al.}(2022){Taverna}, {D{\'\i}az-Gim{\'e}nez}, {Zandivarez}, \& {Mamon}}]{Taverna2022}
{Taverna}, A., {D{\'\i}az-Gim{\'e}nez}, E., {Zandivarez}, A., \& {Mamon}, G.~A. 2022, \mnras, 511, 4741

\bibitem[{{Tricottet} {et~al.}(2025){Tricottet}, {Mamon}, \& {D{\'\i}az-Gim{\'e}nez}}]{Tricottet2025}
{Tricottet}, M., {Mamon}, G.~A., \& {D{\'\i}az-Gim{\'e}nez}, E. 2025, \aap, 699, A329

\bibitem[{{Tzanavaris} {et~al.}(2019){Tzanavaris}, {Gallagher}, {Ali}, {Miller}, {Pentinga}, \& {Johnson}}]{Tza2019}
{Tzanavaris}, P., {Gallagher}, S.~C., {Ali}, S., {et~al.} 2019, \apj, 871, 242

\bibitem[{{Vogelsberger} {et~al.}(2013){Vogelsberger}, {Genel}, {Sijacki}, {Torrey}, {Springel}, \& {Hernquist}}]{Vogelsberger2013}
{Vogelsberger}, M., {Genel}, S., {Sijacki}, D., {et~al.} 2013, \mnras, 436, 3031

\bibitem[{{Vogelsberger} {et~al.}(2014){Vogelsberger}, {Genel}, {Springel}, {Torrey}, {Sijacki}, {Xu}, {Snyder}, {Bird}, {Nelson}, \& {Hernquist}}]{Vogelsberger2014}
{Vogelsberger}, M., {Genel}, S., {Springel}, V., {et~al.} 2014, \nat, 509, 177

\bibitem[{{Walke} \& {Mamon}(1989)}]{WalkeMamon1989}
{Walke}, D.~G. \& {Mamon}, G.~A. 1989, \aap, 225, 291

\bibitem[{{Weinberger} {et~al.}(2017){Weinberger}, {Springel}, {Hernquist}, {Pillepich}, {Marinacci}, {Pakmor}, {Nelson}, {Genel}, {Vogelsberger}, {Naiman}, \& {Torrey}}]{Weinberger2017}
{Weinberger}, R., {Springel}, V., {Hernquist}, L., {et~al.} 2017, \mnras, 465, 3291

\bibitem[{{Wiens} {et~al.}(2019){Wiens}, {Wenger}, {Tzanavaris}, {Johnson}, {Gallagher}, \& {Xiao}}]{Wiens2019}
{Wiens}, C.~D., {Wenger}, T.~V., {Tzanavaris}, P., {et~al.} 2019, \apj, 873, 124

\bibitem[{{Willmer}(2018)}]{Willmer2018}
{Willmer}, C. N.~A. 2018, \apjs, 236, 47

\bibitem[{{York} {et~al.}(2000){York}, {Adelman}, {Anderson}, {Anderson}, {Annis}, {Bahcall}, {Bakken}, {Barkhouser}, {Bastian}, {Berman}, {Boroski}, {Bracker}, {Briegel}, {Briggs}, {Brinkmann}, {Brunner}, {Burles}, {Carey}, {Carr}, {Castander}, {Chen}, {Colestock}, {Connolly}, {Crocker}, {Csabai}, {Czarapata}, {Davis}, {Doi}, {Dombeck}, {Eisenstein}, {Ellman}, {Elms}, {Evans}, {Fan}, {Federwitz}, {Fiscelli}, {Friedman}, {Frieman}, {Fukugita}, {Gillespie}, {Gunn}, {Gurbani}, {de Haas}, {Haldeman}, {Harris}, {Hayes}, {Heckman}, {Hennessy}, {Hindsley}, {Holm}, {Holmgren}, {Huang}, {Hull}, {Husby}, {Ichikawa}, {Ichikawa}, {Ivezi{\'c}}, {Kent}, {Kim}, {Kinney}, {Klaene}, {Kleinman}, {Kleinman}, {Knapp}, {Korienek}, {Kron}, {Kunszt}, {Lamb}, {Lee}, {Leger}, {Limmongkol}, {Lindenmeyer}, {Long}, {Loomis}, {Loveday}, {Lucinio}, {Lupton}, {MacKinnon}, {Mannery}, {Mantsch}, {Margon}, {McGehee}, {McKay}, {Meiksin}, {Merelli}, {Monet}, {Munn}, {Narayanan}, {Nash}, {Neilsen}, {Neswold}, {Newberg}, {Nichol}, {Nicinski},
  {Nonino}, {Okada}, {Okamura}, {Ostriker}, {Owen}, {Pauls}, {Peoples}, {Peterson}, {Petravick}, {Pier}, {Pope}, {Pordes}, {Prosapio}, {Rechenmacher}, {Quinn}, {Richards}, {Richmond}, {Rivetta}, {Rockosi}, {Ruthmansdorfer}, {Sandford}, {Schlegel}, {Schneider}, {Sekiguchi}, {Sergey}, {Shimasaku}, {Siegmund}, {Smee}, {Smith}, {Snedden}, {Stone}, {Stoughton}, {Strauss}, {Stubbs}, {SubbaRao}, {Szalay}, {Szapudi}, {Szokoly}, {Thakar}, {Tremonti}, {Tucker}, {Uomoto}, {Vanden Berk}, {Vogeley}, {Waddell}, {Wang}, {Watanabe}, {Weinberg}, {Yanny}, {Yasuda}, \& {SDSS Collaboration}}]{York2000}
{York}, D.~G., {Adelman}, J., {Anderson}, John~E., J., {et~al.} 2000, \aj, 120, 1579

\bibitem[{{Zandivarez} {et~al.}(2023){Zandivarez}, {D{\'\i}az-Gim{\'e}nez}, {Taverna}, \& {Mamon}}]{Zandivarez2023}
{Zandivarez}, A., {D{\'\i}az-Gim{\'e}nez}, E., {Taverna}, A., \& {Mamon}, G.~A. 2023, \mnras, 526, 3697

\bibitem[{{Zheng} \& {Shen}(2021)}]{Zheng2021}
{Zheng}, Y.-L. \& {Shen}, S.-Y. 2021, \apj, 911, 105

\bibitem[{{Zheng} {et~al.}(2022){Zheng}, {Shen}, \& {Feng}}]{Zheng2022}
{Zheng}, Y.-L., {Shen}, S.-Y., \& {Feng}, S. 2022, \apj, 926, 119

\end{thebibliography}

\begin{appendix}

\section{Properties of compact groups of TNG100 identified at $z=0$}

\begin{table}[!ht]
\caption{Main properties of the 15 CGs identified at redshift $z=0$ in TNG100.}
\label{tab:CGs_z0}
\centering

\begin{tabular}{r c c c c c c}
\hline\hline \\[-10pt]
FoF & $N$ & $\overline{\mu}_r$ & $\overline{R}_{\rm p}$ & $\sigma_{v,\rm 1D}$ & $\log(M_{*})$ & $\log(M_{200})$\\[2pt]
Group & Members & $\left[\mathrm{mag~arcsec^{-2}} \right]$ & $\left[\mathrm{kpc} \right]$ & $\left[ \mathrm{km~s^{-1}} \right]$ & $\left[M_{\odot}\right]$ & $\left[M_{\odot}\right]$\\[3pt]
\hline
24 & 4 & 24.52 & 240 & 386.82 & 12.0 & 13.8 \\[2pt]
57 & 4 & 26.10 & 329 & 226.90 & 11.8 & 13.7 \\[2pt]
61 & 7 & 25.65 & 301 & 319.85 & 11.8 & 13.6 \\[2pt]
142 & 4 & 24.31 & 98 & 193.41 & 11.4 & 13.2 \\[2pt]
230 & 4 & 26.20 & 212 & 186.48 & 11.2 & 12.9 \\[2pt]
231 & 4 & 26.30 & 277 & 142.61 & 11.3 & 12.9 \\[2pt]
239 & 5 & 25.33 & 131 & 207.85 & 11.2 & 12.9 \\[2pt]
243 & 5 & 26.17 & 212 & 192.26 & 11.2 & 12.9 \\[2pt]
263 & 4 & 25.49 & 157 & 136.25 & 11.3 & 12.9 \\[2pt]
302 & 4 & 26.08 & 179 & 115.47 & 11.1 & 12.8 \\[2pt]
416 & 4 & 24.87 & 82 & 149.53 & 10.9 & 12.8 \\[2pt]
424 & 4 & 25.83 & 122 & 161.66 & 11.0 & 12.7 \\[2pt]
430 & 4 & 26.07 & 129 & 103.35 & 10.9 & 12.7 \\[2pt]
469 & 4 & 25.83 & 131 & 109.70 & 11.0 & 12.6 \\[2pt]
485 & 4 & 26.11 & 142 & 144.34 & 10.9 & 12.6 \\[2pt]
\hline
\end{tabular}
\tablefoot{From left to right: FoF Group ID, number of galaxy members, mean surface brightness, mean projected radii, 1D velocity dispersion, total stellar mass and total virial mass. The group 469 is the one shown in Fig. \ref{fig:sph_469}.}
\end{table}
\section{Velocity dispersion of rich groups of TNG100}

\begin{figure}[!h]
    \centering
    \includegraphics[width=0.7\columnwidth]{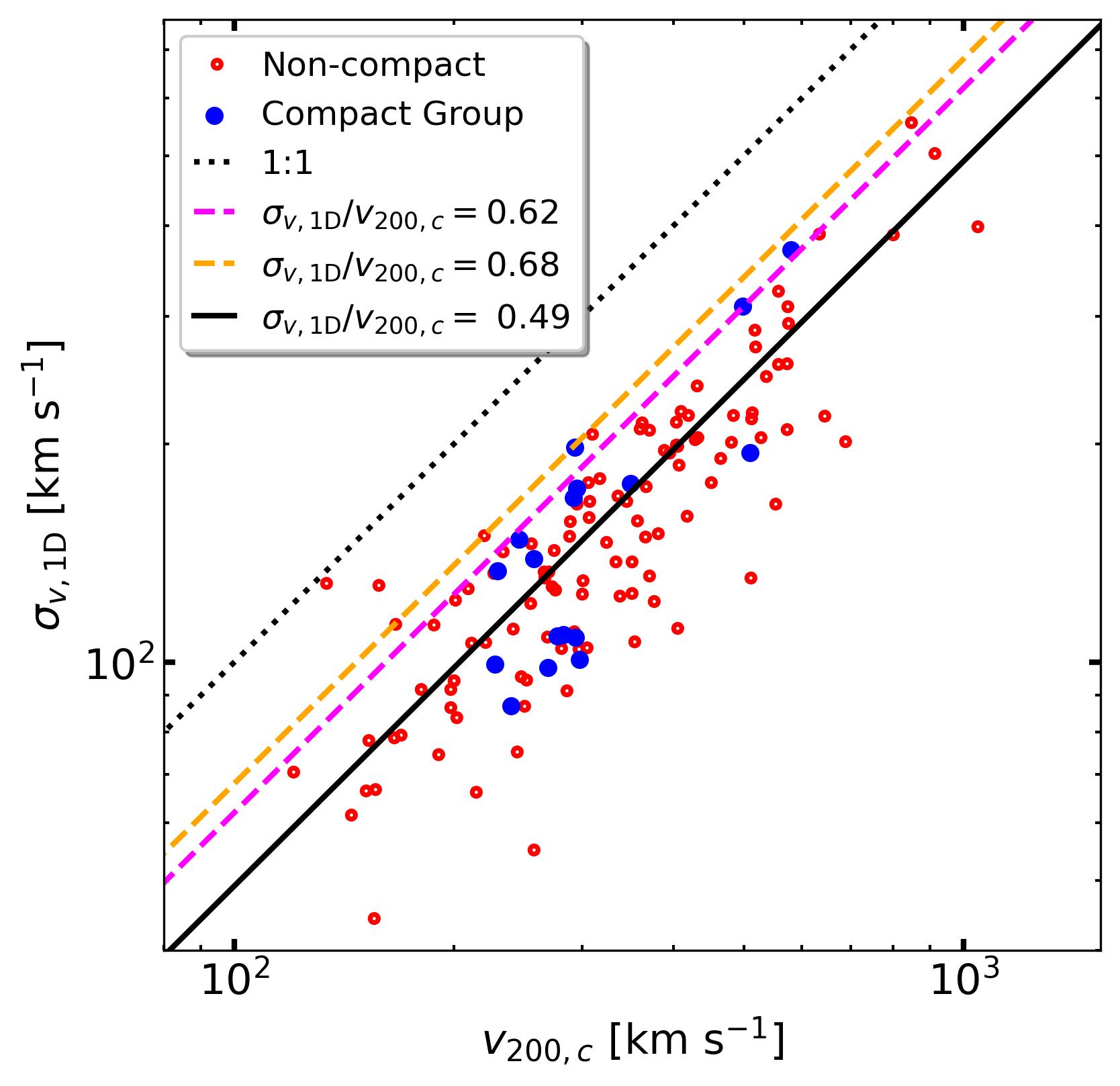}
    \caption{Velocity dispersion as function of virial velocity for rich $(N \geq 4)$ groups of TNG100. We use the same 131 rich groups of Fig. \ref{fig:mu-r_p-sigma-M_T_z0}. Lines show different $\sigma_{v,{\rm 1D}}/v_{200,c}$ ratios, as indicated by the labels.}
    \label{fig:App_sigma_v200}
\end{figure}

As mentioned in Sec. \ref{sec:CG_z0}, we measure the velocity dispersion of rich $(N \geq 4)$ groups (compact or not) and compare it with their virial velocity, as shown in Fig. \ref{fig:App_sigma_v200}. These low-mass groups from TNG100 exhibit a $\sigma_{v,{\rm 1D}}/v_{200,c} \approx 0.5$ (solid line), lower than previously theoretical predictions (dashed lines).

\end{appendix}

\end{document}